\documentclass[aps, prb, twocolumn, preprintnumbers, amsmath, amssymb, superscriptaddress, 10pt]{revtex4-2}
\usepackage{graphicx}
\usepackage{bm}
\usepackage{amssymb}
\usepackage{color}
\usepackage{amsmath}
\usepackage{palatino}	
\usepackage{txfonts} 
\usepackage[colorlinks=true, linkcolor=blue, anchorcolor=black, citecolor=blue, filecolor=black, menucolor=black, runcolor=black, urlcolor=blue]{hyperref}
\usepackage{orcidlink}

\begin{document}
	
\title{Non-Hermitian Cooperation Induced Chern Insulator}

\author{Weiwei Zhu}
\email{phyzhuw@tongji.edu.cn}
\affiliation{Center for Phononics and Thermal Energy Science, China-EU Joint Lab on Nanophononics, Shanghai Key Laboratory of Special Artificial Microstructure Materials and Technology, School of Physics Science and Engineering, Tongji University, Shanghai 200092, China}

\author{Bingbing Wang}
\affiliation{Department of Physics, The Chinese University of Hong Kong, Shatin, Hong Kong SAR, China}

\author{Mengxiang Xie}
\affiliation{Center for Phononics and Thermal Energy Science, China-EU Joint Lab on Nanophononics, Shanghai Key Laboratory of Special Artificial Microstructure Materials and Technology, School of Physics Science and Engineering, Tongji University, Shanghai 200092, China}

\author{Hong-yu Zou}
\affiliation{School of Physics and Electronic Engineering, Jiangsu University, Zhenjiang 212013, China}

\author{Yang Long}
\affiliation{Center for Phononics and Thermal Energy Science, China-EU Joint Lab on Nanophononics, Shanghai Key Laboratory of Special Artificial Microstructure Materials and Technology, School of Physics Science and Engineering, Tongji University, Shanghai 200092, China}

\author{Hong-xiang Sun}
\affiliation{School of Physics and Electronic Engineering, Jiangsu University, Zhenjiang 212013, China}
\affiliation{State Key Laboratory of Acoustics and Marine Information, Institute of Acoustics, Chinese Academy of Sciences, Beijing 100190, China}

\author{Haoran Xue}
\affiliation{Department of Physics, The Chinese University of Hong Kong, Shatin, Hong Kong SAR, China}
\affiliation{State Key Laboratory of Quantum Information Technologies and Materials, The Chinese University of Hong Kong, Shatin, Hong Kong SAR, China}

\author{Jie Ren}
\email{xonics@tongji.edu.cn}
\affiliation{Center for Phononics and Thermal Energy Science, China-EU Joint Lab on Nanophononics, Shanghai Key Laboratory of Special Artificial Microstructure Materials and Technology, School of Physics Science and Engineering, Tongji University, Shanghai 200092, China}

\begin{abstract}
Conventional Chern insulators rely on breaking time-reversal symmetry via magnetic fields or synthetic gauge fields. Here we introduce a fundamentally different paradigm: a static lattice without any gauge field becomes a Chern insulator solely through the cooperation of two non-Hermitian ingredients—on-site loss and non-reciprocal coupling. Neither effect alone suffices; only their synergy opens a topological gap and gives rise to robust chiral edge states with a nonzero Chern number. The underlying mechanism originates from a non-Hermitian chiral molecule formed by the lossy sites, whose dynamically dominant mode carries an orbital angular momentum that effectively generates a synthetic flux after the fast-decaying degrees of freedom are eliminated. Our results establish non-Hermitian cooperation as a versatile resource for inducing Chern topology without external fields or temporal modulations, opening a new route to topological phases in open classical and quantum systems.
\end{abstract}

\maketitle

\textit{Introduction.}—Chern insulators are a paradigmatic class of topological phases, characterized by a quantized Hall conductance and protected chiral edge states that arise from broken time-reversal symmetry~\cite{PhysRevLett.49.405,KOHMOTO1985343,PhysRevLett.61.2015,PhysRevLett.71.3697}. Originally discovered in the quantum Hall effect in two-dimensional electron gases~\cite{PhysRevLett.49.405,KOHMOTO1985343,PhysRevLett.61.2015,PhysRevLett.71.3697}, Chern topology has since transcended electronic systems and found powerful realizations in photonic and acoustic metamaterials~\cite{RN1,RN3,RN2,PhysRevLett.113.087403,RN4,scienceaar4003,scienceaar4005,RevModPhys.91.015006,RN7,PhysRevLett.114.114301,RN5,PhysRevLett.122.014302}. In these classical wave platforms, topological protection provides robustness against disorder and defects, enabling unprecedented control over light and sound propagation with far-reaching applications in topological lasers, non-reciprocal devices, and robust waveguides~\cite{RN3,RN2,PhysRevLett.113.087403,RN4,scienceaar4003,scienceaar4005,RevModPhys.91.015006,RN7,PhysRevLett.114.114301,RN5,PhysRevLett.122.014302}. Conventional approaches to realizing Chern insulators rely on breaking time-reversal symmetry via external magnetic fields, Floquet engineering with periodic driving, or synthetic gauge fields from geometric phases or tailored couplings~\cite{PhysRevB.14.2239,Goldman_2014,RN6,RN8}. These methods, although powerful, are implemented within Hermitian frameworks and incur substantial experimental overhead, particularly in photonic and acoustic systems where magnetic responses are weak and dynamical modulations are challenging~\cite{RN1,RN3,RN2,PhysRevLett.113.087403,RN4,scienceaar4003,scienceaar4005,RevModPhys.91.015006,RN7,PhysRevLett.114.114301,RN5,PhysRevLett.122.014302,PhysRevB.14.2239,Goldman_2014,RN6,RN8}. By contrast, non-Hermitian systems offer a far simpler route to breaking time-reversal symmetry, for example through on-site gain or loss, which are readily realizable. Yet it remains unclear whether these non-Hermitian ingredients alone, without any pre-existing Hermitian topology or external gauge fields, can induce a Chern insulator. This open question is precisely the focus of our work.

The advent of non-Hermitian physics has profoundly enriched the topological landscape~\cite{Ashida02072020,RevModPhys.93.015005,annurevconmatphys,zhang2022review,RN10,RN11}. Open systems with gain, loss, and non-reciprocal couplings exhibit phenomena absent in Hermitian counterparts, including exceptional points where eigenvalues and eigenstates coalesce, complex-energy gaps with novel symmetry classifications, and the non-Hermitian skin effect, a macroscopic accumulation of bulk states at boundaries~\cite{PhysRevX.8.031079,PhysRevLett.120.146402,PhysRevX.9.041015,PhysRevLett.121.086803,PhysRevLett.121.026808,PhysRevLett.124.086801,PhysRevB.99.201103,PhysRevLett.126.010401}. These features have spurred intense research into non-Hermitian topological phases, uncovering point-gap and line-gap topologies, topological knots, and rich bulk-boundary correspondences that go beyond the Hermitian paradigm~\cite{RevModPhys.93.015005,annurevconmatphys,zhang2022review,RN10,RN11,PhysRevX.8.031079,PhysRevLett.120.146402,PhysRevX.9.041015,PhysRevLett.121.086803,PhysRevLett.121.026808,PhysRevLett.124.086801,PhysRevB.99.201103,PhysRevLett.126.010401}.

Despite these advances, the connection between non-Hermiticity and Chern topology remains unsettled~\cite{PhysRevLett.121.136802,PhysRevB.98.165148,PhysRevB.100.155117,PhysRevB.100.081104,PhysRevLett.123.016805,PhysRevB.106.035425,PhysRevB.106.L041302,PhysRevLett.132.113802,PhysRevLett.132.063804,PhysRevLett.133.126601}. Most existing studies of non-Hermitian Chern insulators begin with a Hermitian system that already possesses a nontrivial Chern number; gain, loss, or non-reciprocal couplings are then introduced as perturbations that modify transport properties, giving rise to hybrid skin-topological modes, exceptional edge states, or dislocation-bound states without changing the underlying topological origin~\cite{PhysRevLett.121.136802,PhysRevB.98.165148,PhysRevB.100.155117,PhysRevB.100.081104,PhysRevLett.123.016805,PhysRevB.106.035425,PhysRevB.106.L041302,PhysRevLett.132.113802,PhysRevLett.132.063804,PhysRevLett.133.126601}. In these cases, the Chern topology is inherited from the Hermitian parent rather than generated by non-Hermiticity itself~\cite{PhysRevLett.121.136802,PhysRevB.98.165148,PhysRevB.100.155117,PhysRevB.100.081104,PhysRevLett.123.016805,PhysRevB.106.035425,PhysRevB.106.L041302,PhysRevLett.132.113802,PhysRevLett.132.063804,PhysRevLett.133.126601}. Meanwhile, the non-Hermitian topological classification has been greatly expanded to include point-gap and line-gap topologies~\cite{RevModPhys.93.015005,annurevconmatphys,zhang2022review,RN10,RN11,PhysRevX.8.031079,PhysRevLett.120.146402,PhysRevX.9.041015,PhysRevLett.121.086803,PhysRevLett.121.026808,PhysRevLett.124.086801,PhysRevB.99.201103,PhysRevLett.126.010401}, but whether non-Hermiticity alone, in the absence of any Hermitian topological parent, can induce a Chern insulator remains an open question~\cite{PhysRevLett.121.136802,PhysRevB.98.165148,PhysRevB.100.155117,PhysRevB.100.081104,PhysRevLett.123.016805,PhysRevB.106.035425,PhysRevB.106.L041302,PhysRevLett.132.113802,PhysRevLett.132.063804,PhysRevLett.133.126601}. Moreover, most studies have treated on-site gain/loss and asymmetric (non-reciprocal) couplings in isolation~\cite{PhysRevX.5.021025,RN12,RN13,RN14,RN15,RN16}. The cooperative interplay between these two fundamental forms of non-Hermiticity, on-site dissipation and non-reciprocal hopping, and its potential to generate topology from a trivial Hermitian parent, have remained largely unexplored~\cite{PhysRevX.5.021025,RN12,RN13,RN14,RN15,RN16}. This gap motivates our central question: can the synergy of two distinct non-Hermitian ingredients, neither of which alone suffices, produce a Chern insulator in a static lattice without any gauge field?

\begin{figure*}
	\includegraphics[width=\linewidth]{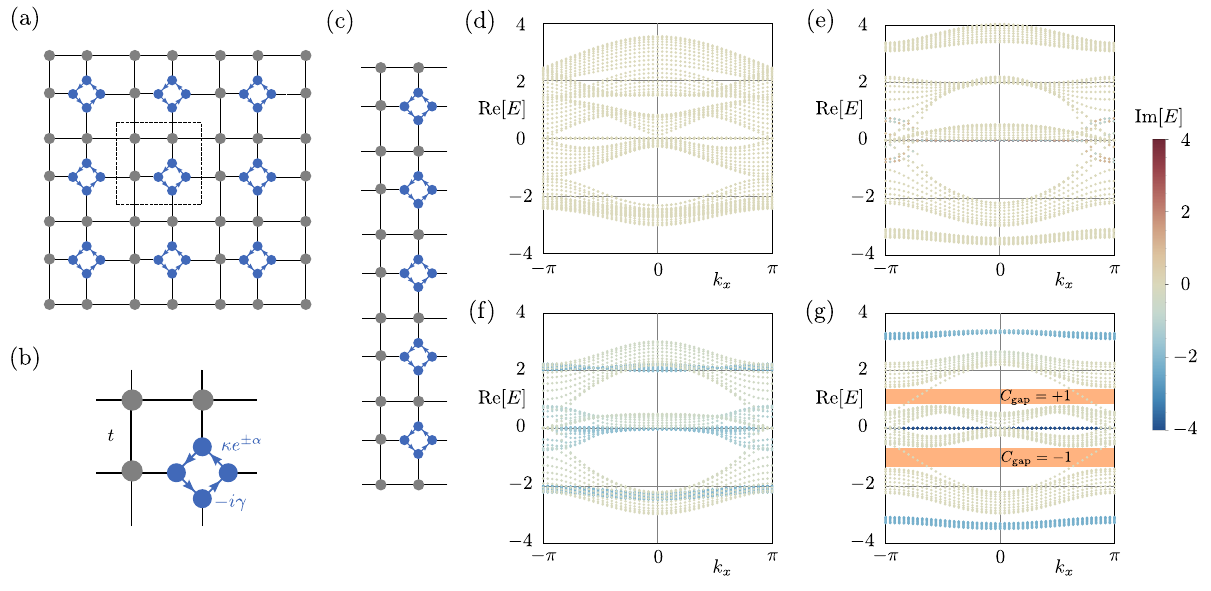}
	\caption{Non-Hermitian cooperation enables a Chern insulator. (a) Schematic of the finite lattice structure. The unit cell is marked by a dashed box, and four lossy sites (blue) form a chiral non-Hermitian "molecule". (b) Enlarged view of a single unit cell, showing the real couplings $t$ (solid black lines), on-site loss $-i\gamma$ (blue dots), and non-reciprocal couplings $\kappa e^{\pm\alpha}$ (blue directed lines, with arrows indicating the hopping direction) (c) Strip geometry used for band calculations, with periodic boundary conditions along $x$ and open boundaries along $y$. (d)--(g) Projected band structures under four conditions: (d) Hermitian limit ($\gamma=0$, $\alpha=0$); (e) only asymmetric couplings ($\gamma=0$, $\alpha=1$); (f) only on-site loss [$\gamma=2\sinh(1)$, $\alpha=0$]; (g) coexistence of both [$\gamma=2\sinh(1)$, $\alpha=1$]. In (g), robust in-gap edge states appear with a nontrivial Chern number $C=1$, and the topological gap is highlighted in orange. All band calculations use $t=1$ and $\kappa=1$, with ten complete unit cells plus one extra row of sites along the $y$ direction.}
	\label{fig1}
\end{figure*}

Here we address this open question by designing a static lattice with seven sites per unit cell, four of which are rendered non-Hermitian through both on-site imaginary potentials and asymmetric (non-reciprocal) inter-site couplings. Crucially, the system contains no artificial gauge fields, magnetic fluxes, or temporal modulations. Our central finding is sharply counterintuitive: neither type of non-Hermiticity alone suffices to open a topological gap. Lattices with on-site loss only, or with asymmetric couplings only, remain trivial. Remarkably, however, when these two non-Hermitian ingredients coexist, the system undergoes a transition into a Chern insulator with a quantized nonzero Chern number. The underlying physical mechanism traces back to the four lossy sites, which form a chiral non-Hermitian "molecule". Its slowest-decaying mode—the one that survives in the long-time dynamics—acquires a chiral phase winding and carries an orbital angular momentum. Through a projection reduction that systematically truncates all fast-decaying degrees of freedom while retaining only this dynamically dominant mode, this emergent orbital texture translates into an effective synthetic flux, mapping our model onto an effective $\pi/2$ flux Hofstadter lattice and thereby revealing the precise origin of the nontrivial topology. The topological protection is further substantiated through numerical analysis of a finite lattice with deliberately introduced defects, where both the eigenenergy spectra and the spatial field profiles unambiguously demonstrate the resilience of the unidirectional edge states against local imperfections. Our work thus establishes a fundamentally new paradigm for realizing Chern topology—not by imposing external fields, synthetic gauge potentials, or pre-existing Hermitian topology, but purely through the cooperation of distinct non-Hermitian effects. This synergy opens an unexplored avenue for engineering topological phases in open classical and quantum systems where Hermitian topological ingredients are otherwise unavailable.

\textit{Model and Results.}—Our lattice consists of seven sites per unit cell, arranged as shown in Fig.~\ref{fig1}(a). Among the seven sites, four are rendered non-Hermitian by two independent ingredients: on-site loss and non-reciprocal inter-site couplings. As illustrated in Fig.~\ref{fig1}(b), each of the four lossy sites carries an imaginary on-site potential $-i\gamma$ (with $\gamma>0$), representing uniform dissipation. Meanwhile, the couplings between adjacent lossy sites are non-reciprocal: a site couples to its clockwise neighbor with amplitude $\kappa e^{+\alpha}$ and to its counterclockwise neighbor with $\kappa e^{-\alpha}$, where $\kappa$ sets the overall coupling strength and $\alpha$ controls the degree of non-reciprocity. The remaining sites are purely Hermitian, with real nearest-neighbor couplings $t$.

In momentum space, the Bloch Hamiltonian takes the block form
\begin{equation}
	H(\mathbf{k}) = 
	\begin{pmatrix}
		H_{\mathrm{Herm}}(\mathbf{k}) & V(\mathbf{k}) \\
		V^\dagger(\mathbf{k}) & H_{\mathrm{NH}}
	\end{pmatrix},
\end{equation}
where $H_{\mathrm{Herm}}$ describes the three Hermitian sites, $H_{\mathrm{NH}}$ is the $4\times4$ non-Hermitian block containing $-i\gamma$ and $\kappa e^{\pm\alpha}$, and $V$ couples the two subspaces. The explicit matrix forms are given in the Supplemental Material.

Importantly, the lattice contains no external magnetic field, no synthetic gauge potential, and no time-dependent modulation; the only sources of non-Hermiticity are the on-site losses and the asymmetric hoppings. To reveal the topological nature of the system, we compute the projected band structures in a strip geometry, periodic along the $x$ direction and open along the $y$ direction, with ten complete unit cells plus one additional row of sites [Fig.~\ref{fig1}(c)]. Throughout the band calculations, we set $t=1$ and $\kappa=1$, and compare four representative parameter regimes: the Hermitian limit ($\gamma=0,\alpha=0$), asymmetric couplings only ($\gamma=0,\alpha=1$), on-site loss only [$\gamma=2\sinh(1),\alpha=0$], and the coexistence of both [$\gamma=2\sinh(1),\alpha=1$]. This parameter choice ensures that the coexistence case operates in the topological regime, as will be demonstrated below.

\begin{figure}
	\includegraphics[width=\linewidth]{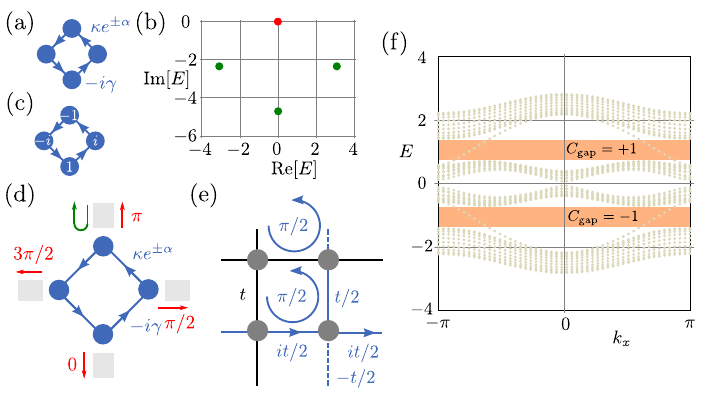}
	\caption{Non-Hermitian chiral molecule, spectral properties, and the effective Hofstadter model. (a) The four lossy sites are extracted from the full lattice to form a non-Hermitian chiral "molecule" with non-reciprocal couplings $\kappa e^{\pm\alpha}$ and on-site loss $-i\gamma$. (b) Complex eigenvalue spectrum of the chiral molecule in the coexistence regime [$\kappa=1$, $\gamma=2\sinh(1)$, $\alpha=1$]. The slowest-decaying mode (red dot, with the smallest imaginary part in magnitude) is well separated from three highly lossy modes (green dots), rendering it dynamically dominant. (c) Eigenstate of the dominant mode, showing a phase winding of $2\pi$ around the four sites, indicating that it carries an orbital angular momentum. (d) Scattering properties of the chiral molecule probed by attaching four ports to the lossy sites. The scattering phase accumulates $\pi/2$ between adjacent ports, yielding a total phase winding of $2\pi$, consistent with the OAM-carrying nature of the dominant mode. (e) Unit cell of the effective Hofstadter model obtained via projection reduction, where the orbital texture of the dominant mode translates into an effective synthetic flux $\pi/2$ through each plaquette. (f) Projected band structure of the effective Hofstadter model in a strip geometry (periodic along $x$, open along $y$), revealing chiral edge states traversing the bulk gap, which reproduces the topological features of the original non-Hermitian lattice.}
	\label{fig2}
\end{figure}

We now examine the projected band structures for the four parameter regimes introduced above, with results summarized in Figs.~\ref{fig1}(d)--(g). In the Hermitian limit ($\gamma=0,\alpha=0$), all couplings are real, and the system satisfies time-reversal symmetry, $T H(\mathbf{k}) T^{-1} = H(-\mathbf{k})$ with $T$ being the complex-conjugation operator. The band structure is purely real and exhibits only partial gaps with no in-gap states, consistent with a trivial phase [Fig.~\ref{fig1}(d)]. When only asymmetric couplings are present ($\gamma=0,\alpha=1$), the system becomes non-Hermitian, yet all hopping amplitudes remain real. Consequently, time-reversal symmetry is preserved, as the non-reciprocal couplings do not introduce any complex phase. The band structure shows partial and full gaps, but no boundary modes emerge [Fig.~\ref{fig1}(e)], and the system remains trivial. Introducing on-site loss alone [$\gamma=2\sinh(1),\alpha=0$] explicitly breaks time-reversal symmetry, since the imaginary on-site potential $-i\gamma$ is incompatible with the condition $T H(\mathbf{k}) T^{-1} = H(-\mathbf{k})$. Nevertheless, despite the loss of time-reversal symmetry, the system does not support any edge state [Fig.~\ref{fig1}(f)], indicating that breaking time-reversal symmetry alone is insufficient to induce a Chern insulator in this lattice. Remarkably, when both ingredients coexist [$\gamma=2\sinh(1),\alpha=1$], time-reversal symmetry is also broken (due to the on-site loss), but now the projected spectrum reveals a pair of robust chiral edge states traversing the bulk band gap, as highlighted in orange in Fig.~\ref{fig1}(g). These edge modes are absent in all other three cases, demonstrating that the nontrivial topology arises solely from the synergy of on-site loss and non-reciprocal coupling. We further confirm the topological nature by computing the Chern number, which yields $C_\text{gap}=\pm1$ for the coexistence regime, consistent with the bulk-boundary correspondence.

\textit{Non-Hermitian Chiral Molecules and Effective Hofstadter Model.}—To understand the physical origin of the topological phase observed in Fig.~\ref{fig1}(g), we analyze the four lossy sites in isolation. As shown in Fig.~\ref{fig2}(a), these sites form a non-Hermitian chiral "molecule" described by the $4\times4$ Hamiltonian $H_{\mathrm{NH}}$, which includes on-site loss $-i\gamma$ and non-reciprocal couplings $\kappa e^{\pm\alpha}$ arranged in a cyclic geometry. Diagonalizing $H_{\mathrm{NH}}$ in the coexistence regime [$\gamma=2\sinh(1), \alpha=1$] reveals a striking spectral structure [Fig.~\ref{fig2}(b)]: among the four complex eigenvalues, one mode (red dot) has an imaginary part substantially smaller in magnitude than the other three (green dots), meaning it decays much more slowly and therefore dominates the long-time dynamics. Crucially, this slowest-decaying mode is not merely a passive survivor; its eigenstate exhibits a phase winding of $2\pi$ around the four sites [Fig.~\ref{fig2}(c)], indicating that it carries an orbital angular momentum (OAM). This OAM-carrying mode serves as the seed for the emergent topology.

The chiral nature of this dominant mode can be further probed through its scattering properties. By attaching four ports to the lossy sites and employing input-output theory, we compute the scattering matrix of the molecule [Fig.~\ref{fig2}(d)]. The scattering phase accumulates by $\pi/2$ between adjacent ports, yielding a total phase winding of $2\pi$ around the loop. This phase winding is a direct scattering signature of the OAM carried by the dominant mode, confirming that the chiral molecule acts as a subwavelength "vortex" that imprints a geometric phase onto transmitted waves. Details are given in the Supplemental Material.

Having identified the OAM-carrying mode as the key ingredient, we now establish a rigorous mapping from the full lattice to an effective model that captures the topology. To this end, we perform a unitary transformation on the full Bloch Hamiltonian $H(\mathbf{k})$ that keeps the Hermitian-sector basis unchanged while diagonalizing the non-Hermitian block $H_{\mathrm{NH}}$. Concretely, we write
\begin{equation}
	\tilde{H}(\mathbf{k}) = U^\dagger H(\mathbf{k}) U,
\end{equation}
where
\begin{equation}
	U = \begin{pmatrix}
		I_3 & 0 \\
		0 & W
	\end{pmatrix},
\end{equation}
with $I_3$ being the $3\times3$ identity matrix acting on the Hermitian sites and $W$ being the $4\times4$ unitary matrix that diagonalizes $H_{\mathrm{NH}}$, i.e., $W^\dagger H_{\mathrm{NH}} W = \mathrm{diag}(\lambda_1,\lambda_2,\lambda_3,\lambda_4)$. In this basis, the four non-Hermitian modes are decoupled, and the coupling between the Hermitian environment and each non-Hermitian mode is explicitly revealed. Since the OAM-carrying mode $\lambda_1$ has the smallest decay rate, it dominates the long-time dynamics. We therefore project the full system onto the subspace of this single dominant mode, explicitly truncating the other three fast-decaying modes. This projection yields an effective Hamiltonian defined on the reduced Hilbert space. Remarkably, the orbital texture of the retained OAM mode translates into an effective synthetic flux threading each plaquette of the reduced lattice, mapping the system exactly onto a Hofstadter model (Details are given in the Supplemental Material.). The unit cell of this effective model is depicted in Fig.~\ref{fig2}(e), where the non-reciprocal phase accumulation mimics a uniform magnetic field.

\begin{figure}
	\includegraphics[width=\linewidth]{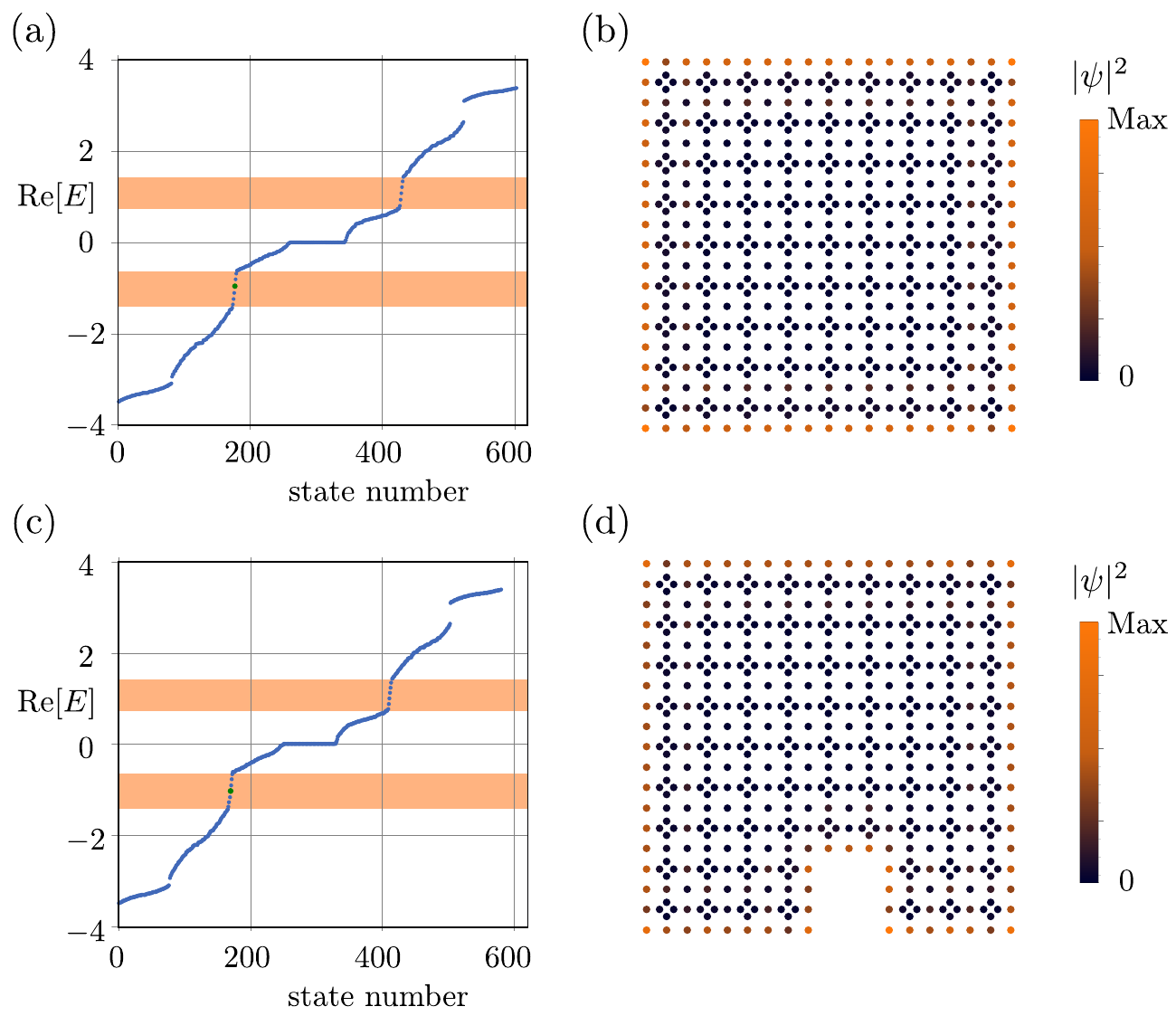}
	\caption{Finite-lattice spectra and field profiles validating the topological protection against local perturbations. (a) Eigenenergy spectrum of a finite $10\times10$ lattice with open boundary conditions; a representative edge state is marked by a green dot. (b) Spatial distribution of the field amplitude for the green-marked edge state, illustrating its boundary localization. (c) Eigenenergy spectrum of the same lattice but with intentionally introduced defects; the persistent edge state (green dot) remains inside the bulk gap. (d) Field profile of the edge state in the defective lattice, demonstrating the robustness of topological edge state. }
	\label{fig3}
\end{figure}

To verify the validity of this reduction, we compute the projected band structure of the effective Hofstadter model in the same strip geometry used for the original lattice. As shown in Fig.~\ref{fig2}(f), the effective model reproduces the low-loss sector of the original spectrum with remarkable fidelity: the bulk bands and, crucially, the chiral edge states traversing the band gap are faithfully captured. The high-loss bulk bands of the original system [Fig.~\ref{fig1}(g)] find their natural correspondence in the modes that have been traced out—they are precisely the three fast-decaying modes of the chiral molecule that were eliminated. This one-to-one correspondence confirms that the reduction correctly separates the dynamically dominant degree of freedom from the short-lived ones, while preserving the essential topology. The excellent agreement between Fig.~\ref{fig2}(f) and the low-loss part of Fig.~\ref{fig1}(g) demonstrates that the effective Hofstadter model captures the full topological content of the original non-Hermitian lattice. This analysis establishes a clear physical picture: the synergy of on-site loss and non-reciprocal coupling creates an OAM-bearing dominant mode, whose phase winding serves as an effective gauge field after projection, thereby generating a Chern insulator without any external magnetic field or synthetic gauge potential.

\textit{Robustness of Topological Edge State.}—To substantiate the topological protection, we analyze a finite \(10\times10\) lattice with open boundaries. In Fig.~3(a), the bulk bands are separated by a spectral gap (shaded in orange), within which gapless topological edge states are clearly observed. One representative edge state is marked by a green dot, and its field profile, displayed in Fig.~3(b), confirms strong localization along the boundary. To test the resilience against local imperfections, we deliberately remove a portion of the bottom boundary to introduce a geometric defect. Remarkably, as shown in Fig.~3(c), the gapless edge states persist inside the gap (one representative highlighted by a green dot), and its corresponding field distribution [Fig.~3(d)] remains extended along the remaining edges, rather than scattering into the bulk. These results unambiguously demonstrate that the edge states are robust against finite local perturbations, corroborating that the non-Hermitian cooperation induces a genuine topological phase with Chern-number protection.

\textit{Conclusion.}—In summary, we have demonstrated that a static lattice devoid of magnetic fields, synthetic gauge potentials, or time-dependent modulations can host a Chern insulator provided that two distinct forms of non-Hermiticity—on-site loss and non-reciprocal couplings—are simultaneously present. Neither ingredient alone suffices; the Hermitian limit, the case with only asymmetric couplings, and the case with only on-site loss all remain trivial. Only their synergy opens a topological gap and produces robust chiral edge states with a nonzero Chern number $C_\text{gap}=\pm1$. The underlying mechanism is traced to the four lossy sites, which constitute a non-Hermitian chiral molecule whose slowest-decaying mode carries an orbital angular momentum. By projecting the system onto the subspace of the dynamically dominant mode and truncating the three fast-decaying ones, the orbital texture of the retained mode manifests as a synthetic magnetic flux, mapping the system onto an effective Hofstadter model. This mapping not only provides a transparent physical interpretation of the emergent topology but also establishes a rigorous theoretical framework that can be generalized to other non-Hermitian platforms.

Our work establishes a new paradigm for engineering Chern topology through the \emph{cooperation} of non-Hermitian effects, as opposed to the conventional routes that rely on external fields or pre-existing Hermitian topology~\cite{PhysRevB.14.2239,Goldman_2014,RN6,RN8,PhysRevLett.121.136802,PhysRevB.98.165148,PhysRevB.100.155117,PhysRevB.100.081104,PhysRevB.106.035425,PhysRevB.106.L041302,PhysRevLett.132.113802,PhysRevLett.132.063804,PhysRevLett.133.126601}. This mechanism can be realized in acoustic, optical, mechanical, and electrical circuits, where gain, loss, and non-reciprocal elements are increasingly accessible~\cite{RN1,RN3,RN2,PhysRevLett.113.087403,RN4,scienceaar4003,scienceaar4005,RevModPhys.91.015006,RN7,PhysRevLett.114.114301,RN5,PhysRevLett.122.014302,PhysRevX.5.021025,RN12,RN13,RN14,RN15,RN16,pnas2010580117,5608038,RN17}. More broadly, our findings suggest that non-Hermitian synergy may serve as a powerful and general principle for inducing topological order in open classical and quantum systems, opening avenues for the discovery of previously inaccessible topological phases and their applications in robust wave manipulation and non-reciprocal device design~\cite{Ashida02072020,RevModPhys.93.015005,annurevconmatphys,zhang2022review,RN10,RN11}.

\bigskip\noindent{\large\textbf{Data availability}}

\noindent All data are available from the corresponding authors upon reasonable request.

\twocolumngrid
\bigskip\noindent{\large\textbf{References}}
\makeatletter
\renewcommand{\bibsection}{}
\makeatother

\bigskip\noindent{\large{\bf{Acknowledgements}}}

\noindent W.Z. is supported by the National Natural Science Foundation of China (Grants No.~12404499). H.S. is supported by the National Natural Science Foundation of China (Grants No.~12274183).  H.X. is supported by NSFC Grant No 62401491 and Research Grants Council Hong Kong SAR, Grant Nos. 24304825 and 14303326. J.R. is supported by the National Key R$\&$D Program of China (2022YFA1404400 and Grant Nos. 2023YFA1406900)

\bigskip\noindent{\large{\bf{Author contributions}}}

\noindent W.Z. initiated the project, performed the numerical simulation and wrote the manuscript with input from all authors.

\bigskip\noindent{\large{\bf{Competing interests}}}

\noindent The authors declare no competing interests.

\bigskip\noindent{\large{\bf{Additional information}}}

\noindent \textbf{Supplementary information} The online version contains supplementary material available at (URL to be inserted upon publication).
	

\begin{thebibliography}{53}%
	\makeatletter
	\providecommand \@ifxundefined [1]{%
		\@ifx{#1\undefined}
	}%
	\providecommand \@ifnum [1]{%
		\ifnum #1\expandafter \@firstoftwo
		\else \expandafter \@secondoftwo
		\fi
	}%
	\providecommand \@ifx [1]{%
		\ifx #1\expandafter \@firstoftwo
		\else \expandafter \@secondoftwo
		\fi
	}%
	\providecommand \natexlab [1]{#1}%
	\providecommand \enquote  [1]{``#1''}%
	\providecommand \bibnamefont  [1]{#1}%
	\providecommand \bibfnamefont [1]{#1}%
	\providecommand \citenamefont [1]{#1}%
	\providecommand \href@noop [0]{\@secondoftwo}%
	\providecommand \href [0]{\begingroup \@sanitize@url \@href}%
	\providecommand \@href[1]{\@@startlink{#1}\@@href}%
	\providecommand \@@href[1]{\endgroup#1\@@endlink}%
	\providecommand \@sanitize@url [0]{\catcode `\\12\catcode `\$12\catcode
		`\&12\catcode `\#12\catcode `\^12\catcode `\_12\catcode `\%12\relax}%
	\providecommand \@@startlink[1]{}%
	\providecommand \@@endlink[0]{}%
	\providecommand \url  [0]{\begingroup\@sanitize@url \@url }%
	\providecommand \@url [1]{\endgroup\@href {#1}{\urlprefix }}%
	\providecommand \urlprefix  [0]{URL }%
	\providecommand \Eprint [0]{\href }%
	\providecommand \doibase [0]{https://doi.org/}%
	\providecommand \selectlanguage [0]{\@gobble}%
	\providecommand \bibinfo  [0]{\@secondoftwo}%
	\providecommand \bibfield  [0]{\@secondoftwo}%
	\providecommand \translation [1]{[#1]}%
	\providecommand \BibitemOpen [0]{}%
	\providecommand \bibitemStop [0]{}%
	\providecommand \bibitemNoStop [0]{.\EOS\space}%
	\providecommand \EOS [0]{\spacefactor3000\relax}%
	\providecommand \BibitemShut  [1]{\csname bibitem#1\endcsname}%
	\let\auto@bib@innerbib\@empty
	\bibitem [{\citenamefont {Thouless}\ \emph {et~al.}(1982)\citenamefont
		{Thouless}, \citenamefont {Kohmoto}, \citenamefont {Nightingale},\ and\
		\citenamefont {den Nijs}}]{PhysRevLett.49.405}%
	\BibitemOpen
	\bibfield  {author} {\bibinfo {author} {\bibfnamefont {D.~J.}\ \bibnamefont
			{Thouless}}, \bibinfo {author} {\bibfnamefont {M.}~\bibnamefont {Kohmoto}},
		\bibinfo {author} {\bibfnamefont {M.~P.}\ \bibnamefont {Nightingale}},\ and\
		\bibinfo {author} {\bibfnamefont {M.}~\bibnamefont {den Nijs}},\ }\bibfield
	{title} {\bibinfo {title} {Quantized hall conductance in a two-dimensional
			periodic potential},\ }\href {https://doi.org/10.1103/PhysRevLett.49.405}
	{\bibfield  {journal} {\bibinfo  {journal} {Phys. Rev. Lett.}\ }\textbf
		{\bibinfo {volume} {49}},\ \bibinfo {pages} {405} (\bibinfo {year}
		{1982})}\BibitemShut {NoStop}%
	\bibitem [{\citenamefont {Kohmoto}(1985)}]{KOHMOTO1985343}%
	\BibitemOpen
	\bibfield  {author} {\bibinfo {author} {\bibfnamefont {M.}~\bibnamefont
			{Kohmoto}},\ }\bibfield  {title} {\bibinfo {title} {Topological invariant and
			the quantization of the hall conductance},\ }\href
	{https://doi.org/https://doi.org/10.1016/0003-4916(85)90148-4} {\bibfield
		{journal} {\bibinfo  {journal} {Annals of Physics}\ }\textbf {\bibinfo
			{volume} {160}},\ \bibinfo {pages} {343} (\bibinfo {year}
		{1985})}\BibitemShut {NoStop}%
	\bibitem [{\citenamefont {Haldane}(1988)}]{PhysRevLett.61.2015}%
	\BibitemOpen
	\bibfield  {author} {\bibinfo {author} {\bibfnamefont {F.~D.~M.}\
			\bibnamefont {Haldane}},\ }\bibfield  {title} {\bibinfo {title} {Model for a
			quantum hall effect without landau levels: Condensed-matter realization of
			the "parity anomaly"},\ }\href {https://doi.org/10.1103/PhysRevLett.61.2015}
	{\bibfield  {journal} {\bibinfo  {journal} {Phys. Rev. Lett.}\ }\textbf
		{\bibinfo {volume} {61}},\ \bibinfo {pages} {2015} (\bibinfo {year}
		{1988})}\BibitemShut {NoStop}%
	\bibitem [{\citenamefont {Hatsugai}(1993)}]{PhysRevLett.71.3697}%
	\BibitemOpen
	\bibfield  {author} {\bibinfo {author} {\bibfnamefont {Y.}~\bibnamefont
			{Hatsugai}},\ }\bibfield  {title} {\bibinfo {title} {Chern number and edge
			states in the integer quantum hall effect},\ }\href
	{https://doi.org/10.1103/PhysRevLett.71.3697} {\bibfield  {journal} {\bibinfo
			{journal} {Phys. Rev. Lett.}\ }\textbf {\bibinfo {volume} {71}},\ \bibinfo
		{pages} {3697} (\bibinfo {year} {1993})}\BibitemShut {NoStop}%
	\bibitem [{\citenamefont {Wang}\ \emph {et~al.}(2009)\citenamefont {Wang},
		\citenamefont {Chong}, \citenamefont {Joannopoulos},\ and\ \citenamefont
		{Soljačić}}]{RN1}%
	\BibitemOpen
	\bibfield  {author} {\bibinfo {author} {\bibfnamefont {Z.}~\bibnamefont
			{Wang}}, \bibinfo {author} {\bibfnamefont {Y.}~\bibnamefont {Chong}},
		\bibinfo {author} {\bibfnamefont {J.~D.}\ \bibnamefont {Joannopoulos}},\ and\
		\bibinfo {author} {\bibfnamefont {M.}~\bibnamefont {Soljačić}},\ }\bibfield
	{title} {\bibinfo {title} {Observation of unidirectional
			backscattering-immune topological electromagnetic states},\ }\href
	{https://doi.org/10.1038/nature08293} {\bibfield  {journal} {\bibinfo
			{journal} {Nature}\ }\textbf {\bibinfo {volume} {461}},\ \bibinfo {pages}
		{772–775} (\bibinfo {year} {2009})}\BibitemShut {NoStop}%
	\bibitem [{\citenamefont {Hafezi}\ \emph {et~al.}(2011)\citenamefont {Hafezi},
		\citenamefont {Demler}, \citenamefont {Lukin},\ and\ \citenamefont
		{Taylor}}]{RN3}%
	\BibitemOpen
	\bibfield  {author} {\bibinfo {author} {\bibfnamefont {M.}~\bibnamefont
			{Hafezi}}, \bibinfo {author} {\bibfnamefont {E.~A.}\ \bibnamefont {Demler}},
		\bibinfo {author} {\bibfnamefont {M.~D.}\ \bibnamefont {Lukin}},\ and\
		\bibinfo {author} {\bibfnamefont {J.~M.}\ \bibnamefont {Taylor}},\ }\bibfield
	{title} {\bibinfo {title} {Robust optical delay lines with topological
			protection},\ }\href {https://doi.org/10.1038/nphys2063} {\bibfield
		{journal} {\bibinfo  {journal} {Nature Physics}\ }\textbf {\bibinfo {volume}
			{7}},\ \bibinfo {pages} {907–912} (\bibinfo {year} {2011})}\BibitemShut
	{NoStop}%
	\bibitem [{\citenamefont {Rechtsman}\ \emph {et~al.}(2013)\citenamefont
		{Rechtsman}, \citenamefont {Zeuner}, \citenamefont {Plotnik}, \citenamefont
		{Lumer}, \citenamefont {Podolsky}, \citenamefont {Dreisow}, \citenamefont
		{Nolte}, \citenamefont {Segev},\ and\ \citenamefont {Szameit}}]{RN2}%
	\BibitemOpen
	\bibfield  {author} {\bibinfo {author} {\bibfnamefont {M.~C.}\ \bibnamefont
			{Rechtsman}}, \bibinfo {author} {\bibfnamefont {J.~M.}\ \bibnamefont
			{Zeuner}}, \bibinfo {author} {\bibfnamefont {Y.}~\bibnamefont {Plotnik}},
		\bibinfo {author} {\bibfnamefont {Y.}~\bibnamefont {Lumer}}, \bibinfo
		{author} {\bibfnamefont {D.}~\bibnamefont {Podolsky}}, \bibinfo {author}
		{\bibfnamefont {F.}~\bibnamefont {Dreisow}}, \bibinfo {author} {\bibfnamefont
			{S.}~\bibnamefont {Nolte}}, \bibinfo {author} {\bibfnamefont
			{M.}~\bibnamefont {Segev}},\ and\ \bibinfo {author} {\bibfnamefont
			{A.}~\bibnamefont {Szameit}},\ }\bibfield  {title} {\bibinfo {title}
		{Photonic floquet topological insulators},\ }\href
	{https://doi.org/10.1038/nature12066} {\bibfield  {journal} {\bibinfo
			{journal} {Nature}\ }\textbf {\bibinfo {volume} {496}},\ \bibinfo {pages}
		{196–200} (\bibinfo {year} {2013})}\BibitemShut {NoStop}%
	\bibitem [{\citenamefont {Mittal}\ \emph {et~al.}(2014)\citenamefont {Mittal},
		\citenamefont {Fan}, \citenamefont {Faez}, \citenamefont {Migdall},
		\citenamefont {Taylor},\ and\ \citenamefont
		{Hafezi}}]{PhysRevLett.113.087403}%
	\BibitemOpen
	\bibfield  {author} {\bibinfo {author} {\bibfnamefont {S.}~\bibnamefont
			{Mittal}}, \bibinfo {author} {\bibfnamefont {J.}~\bibnamefont {Fan}},
		\bibinfo {author} {\bibfnamefont {S.}~\bibnamefont {Faez}}, \bibinfo {author}
		{\bibfnamefont {A.}~\bibnamefont {Migdall}}, \bibinfo {author} {\bibfnamefont
			{J.~M.}\ \bibnamefont {Taylor}},\ and\ \bibinfo {author} {\bibfnamefont
			{M.}~\bibnamefont {Hafezi}},\ }\bibfield  {title} {\bibinfo {title}
		{Topologically robust transport of photons in a synthetic gauge field},\
	}\href {https://doi.org/10.1103/PhysRevLett.113.087403} {\bibfield  {journal}
		{\bibinfo  {journal} {Phys. Rev. Lett.}\ }\textbf {\bibinfo {volume} {113}},\
		\bibinfo {pages} {087403} (\bibinfo {year} {2014})}\BibitemShut {NoStop}%
	\bibitem [{\citenamefont {Lu}\ \emph {et~al.}(2014)\citenamefont {Lu},
		\citenamefont {Joannopoulos},\ and\ \citenamefont {Soljačić}}]{RN4}%
	\BibitemOpen
	\bibfield  {author} {\bibinfo {author} {\bibfnamefont {L.}~\bibnamefont
			{Lu}}, \bibinfo {author} {\bibfnamefont {J.~D.}\ \bibnamefont
			{Joannopoulos}},\ and\ \bibinfo {author} {\bibfnamefont {M.}~\bibnamefont
			{Soljačić}},\ }\bibfield  {title} {\bibinfo {title} {Topological
			photonics},\ }\href {https://doi.org/10.1038/nphoton.2014.248} {\bibfield
		{journal} {\bibinfo  {journal} {Nature Photonics}\ }\textbf {\bibinfo
			{volume} {8}},\ \bibinfo {pages} {821–829} (\bibinfo {year}
		{2014})}\BibitemShut {NoStop}%
	\bibitem [{\citenamefont {Harari}\ \emph {et~al.}(2018)\citenamefont {Harari},
		\citenamefont {Bandres}, \citenamefont {Lumer}, \citenamefont {Rechtsman},
		\citenamefont {Chong}, \citenamefont {Khajavikhan}, \citenamefont
		{Christodoulides},\ and\ \citenamefont {Segev}}]{scienceaar4003}%
	\BibitemOpen
	\bibfield  {author} {\bibinfo {author} {\bibfnamefont {G.}~\bibnamefont
			{Harari}}, \bibinfo {author} {\bibfnamefont {M.~A.}\ \bibnamefont {Bandres}},
		\bibinfo {author} {\bibfnamefont {Y.}~\bibnamefont {Lumer}}, \bibinfo
		{author} {\bibfnamefont {M.~C.}\ \bibnamefont {Rechtsman}}, \bibinfo {author}
		{\bibfnamefont {Y.~D.}\ \bibnamefont {Chong}}, \bibinfo {author}
		{\bibfnamefont {M.}~\bibnamefont {Khajavikhan}}, \bibinfo {author}
		{\bibfnamefont {D.~N.}\ \bibnamefont {Christodoulides}},\ and\ \bibinfo
		{author} {\bibfnamefont {M.}~\bibnamefont {Segev}},\ }\bibfield  {title}
	{\bibinfo {title} {Topological insulator laser: Theory},\ }\href
	{https://doi.org/10.1126/science.aar4003} {\bibfield  {journal} {\bibinfo
			{journal} {Science}\ }\textbf {\bibinfo {volume} {359}},\ \bibinfo {pages}
		{eaar4003} (\bibinfo {year} {2018})}\BibitemShut {NoStop}%
	\bibitem [{\citenamefont {Bandres}\ \emph {et~al.}(2018)\citenamefont
		{Bandres}, \citenamefont {Wittek}, \citenamefont {Harari}, \citenamefont
		{Parto}, \citenamefont {Ren}, \citenamefont {Segev}, \citenamefont
		{Christodoulides},\ and\ \citenamefont {Khajavikhan}}]{scienceaar4005}%
	\BibitemOpen
	\bibfield  {author} {\bibinfo {author} {\bibfnamefont {M.~A.}\ \bibnamefont
			{Bandres}}, \bibinfo {author} {\bibfnamefont {S.}~\bibnamefont {Wittek}},
		\bibinfo {author} {\bibfnamefont {G.}~\bibnamefont {Harari}}, \bibinfo
		{author} {\bibfnamefont {M.}~\bibnamefont {Parto}}, \bibinfo {author}
		{\bibfnamefont {J.}~\bibnamefont {Ren}}, \bibinfo {author} {\bibfnamefont
			{M.}~\bibnamefont {Segev}}, \bibinfo {author} {\bibfnamefont {D.~N.}\
			\bibnamefont {Christodoulides}},\ and\ \bibinfo {author} {\bibfnamefont
			{M.}~\bibnamefont {Khajavikhan}},\ }\bibfield  {title} {\bibinfo {title}
		{Topological insulator laser: Experiments},\ }\href
	{https://doi.org/10.1126/science.aar4005} {\bibfield  {journal} {\bibinfo
			{journal} {Science}\ }\textbf {\bibinfo {volume} {359}},\ \bibinfo {pages}
		{eaar4005} (\bibinfo {year} {2018})}\BibitemShut {NoStop}%
	\bibitem [{\citenamefont {Ozawa}\ \emph {et~al.}(2019)\citenamefont {Ozawa},
		\citenamefont {Price}, \citenamefont {Amo}, \citenamefont {Goldman},
		\citenamefont {Hafezi}, \citenamefont {Lu}, \citenamefont {Rechtsman},
		\citenamefont {Schuster}, \citenamefont {Simon}, \citenamefont {Zilberberg},\
		and\ \citenamefont {Carusotto}}]{RevModPhys.91.015006}%
	\BibitemOpen
	\bibfield  {author} {\bibinfo {author} {\bibfnamefont {T.}~\bibnamefont
			{Ozawa}}, \bibinfo {author} {\bibfnamefont {H.~M.}\ \bibnamefont {Price}},
		\bibinfo {author} {\bibfnamefont {A.}~\bibnamefont {Amo}}, \bibinfo {author}
		{\bibfnamefont {N.}~\bibnamefont {Goldman}}, \bibinfo {author} {\bibfnamefont
			{M.}~\bibnamefont {Hafezi}}, \bibinfo {author} {\bibfnamefont
			{L.}~\bibnamefont {Lu}}, \bibinfo {author} {\bibfnamefont {M.~C.}\
			\bibnamefont {Rechtsman}}, \bibinfo {author} {\bibfnamefont {D.}~\bibnamefont
			{Schuster}}, \bibinfo {author} {\bibfnamefont {J.}~\bibnamefont {Simon}},
		\bibinfo {author} {\bibfnamefont {O.}~\bibnamefont {Zilberberg}},\ and\
		\bibinfo {author} {\bibfnamefont {I.}~\bibnamefont {Carusotto}},\ }\bibfield
	{title} {\bibinfo {title} {Topological photonics},\ }\href
	{https://doi.org/10.1103/RevModPhys.91.015006} {\bibfield  {journal}
		{\bibinfo  {journal} {Rev. Mod. Phys.}\ }\textbf {\bibinfo {volume} {91}},\
		\bibinfo {pages} {015006} (\bibinfo {year} {2019})}\BibitemShut {NoStop}%
	\bibitem [{\citenamefont {Yan}\ \emph {et~al.}(2024)\citenamefont {Yan},
		\citenamefont {Zhang},\ and\ \citenamefont {Chen}}]{RN7}%
	\BibitemOpen
	\bibfield  {author} {\bibinfo {author} {\bibfnamefont {W.}~\bibnamefont
			{Yan}}, \bibinfo {author} {\bibfnamefont {B.}~\bibnamefont {Zhang}},\ and\
		\bibinfo {author} {\bibfnamefont {F.}~\bibnamefont {Chen}},\ }\bibfield
	{title} {\bibinfo {title} {Photonic topological insulators in femtosecond
			laser direct-written waveguides},\ }\href
	{https://doi.org/10.1038/s44310-024-00040-7} {\bibfield  {journal} {\bibinfo
			{journal} {npj Nanophotonics}\ }\textbf {\bibinfo {volume} {1}},\ \bibinfo
		{pages} {40} (\bibinfo {year} {2024})}\BibitemShut {NoStop}%
	\bibitem [{\citenamefont {Yang}\ \emph {et~al.}(2015)\citenamefont {Yang},
		\citenamefont {Gao}, \citenamefont {Shi}, \citenamefont {Lin}, \citenamefont
		{Gao}, \citenamefont {Chong},\ and\ \citenamefont
		{Zhang}}]{PhysRevLett.114.114301}%
	\BibitemOpen
	\bibfield  {author} {\bibinfo {author} {\bibfnamefont {Z.}~\bibnamefont
			{Yang}}, \bibinfo {author} {\bibfnamefont {F.}~\bibnamefont {Gao}}, \bibinfo
		{author} {\bibfnamefont {X.}~\bibnamefont {Shi}}, \bibinfo {author}
		{\bibfnamefont {X.}~\bibnamefont {Lin}}, \bibinfo {author} {\bibfnamefont
			{Z.}~\bibnamefont {Gao}}, \bibinfo {author} {\bibfnamefont {Y.}~\bibnamefont
			{Chong}},\ and\ \bibinfo {author} {\bibfnamefont {B.}~\bibnamefont {Zhang}},\
	}\bibfield  {title} {\bibinfo {title} {Topological acoustics},\ }\href
	{https://doi.org/10.1103/PhysRevLett.114.114301} {\bibfield  {journal}
		{\bibinfo  {journal} {Phys. Rev. Lett.}\ }\textbf {\bibinfo {volume} {114}},\
		\bibinfo {pages} {114301} (\bibinfo {year} {2015})}\BibitemShut {NoStop}%
	\bibitem [{\citenamefont {He}\ \emph {et~al.}(2016)\citenamefont {He},
		\citenamefont {Ni}, \citenamefont {Ge}, \citenamefont {Sun}, \citenamefont
		{Chen}, \citenamefont {Lu}, \citenamefont {Liu},\ and\ \citenamefont
		{Chen}}]{RN5}%
	\BibitemOpen
	\bibfield  {author} {\bibinfo {author} {\bibfnamefont {C.}~\bibnamefont
			{He}}, \bibinfo {author} {\bibfnamefont {X.}~\bibnamefont {Ni}}, \bibinfo
		{author} {\bibfnamefont {H.}~\bibnamefont {Ge}}, \bibinfo {author}
		{\bibfnamefont {X.-C.}\ \bibnamefont {Sun}}, \bibinfo {author} {\bibfnamefont
			{Y.-B.}\ \bibnamefont {Chen}}, \bibinfo {author} {\bibfnamefont {M.-H.}\
			\bibnamefont {Lu}}, \bibinfo {author} {\bibfnamefont {X.-P.}\ \bibnamefont
			{Liu}},\ and\ \bibinfo {author} {\bibfnamefont {Y.-F.}\ \bibnamefont
			{Chen}},\ }\bibfield  {title} {\bibinfo {title} {Acoustic topological
			insulator and robust one-way sound transport},\ }\href
	{https://doi.org/10.1038/nphys3867} {\bibfield  {journal} {\bibinfo
			{journal} {Nature Physics}\ }\textbf {\bibinfo {volume} {12}},\ \bibinfo
		{pages} {1124–1129} (\bibinfo {year} {2016})}\BibitemShut {NoStop}%
	\bibitem [{\citenamefont {Ding}\ \emph {et~al.}(2019)\citenamefont {Ding},
		\citenamefont {Peng}, \citenamefont {Zhu}, \citenamefont {Fan}, \citenamefont
		{Yang}, \citenamefont {Liang}, \citenamefont {Zhu}, \citenamefont {Wan},\
		and\ \citenamefont {Cheng}}]{PhysRevLett.122.014302}%
	\BibitemOpen
	\bibfield  {author} {\bibinfo {author} {\bibfnamefont {Y.}~\bibnamefont
			{Ding}}, \bibinfo {author} {\bibfnamefont {Y.}~\bibnamefont {Peng}}, \bibinfo
		{author} {\bibfnamefont {Y.}~\bibnamefont {Zhu}}, \bibinfo {author}
		{\bibfnamefont {X.}~\bibnamefont {Fan}}, \bibinfo {author} {\bibfnamefont
			{J.}~\bibnamefont {Yang}}, \bibinfo {author} {\bibfnamefont {B.}~\bibnamefont
			{Liang}}, \bibinfo {author} {\bibfnamefont {X.}~\bibnamefont {Zhu}}, \bibinfo
		{author} {\bibfnamefont {X.}~\bibnamefont {Wan}},\ and\ \bibinfo {author}
		{\bibfnamefont {J.}~\bibnamefont {Cheng}},\ }\bibfield  {title} {\bibinfo
		{title} {Experimental demonstration of acoustic chern insulators},\ }\href
	{https://doi.org/10.1103/PhysRevLett.122.014302} {\bibfield  {journal}
		{\bibinfo  {journal} {Phys. Rev. Lett.}\ }\textbf {\bibinfo {volume} {122}},\
		\bibinfo {pages} {014302} (\bibinfo {year} {2019})}\BibitemShut {NoStop}%
	\bibitem [{\citenamefont {Hofstadter}(1976)}]{PhysRevB.14.2239}%
	\BibitemOpen
	\bibfield  {author} {\bibinfo {author} {\bibfnamefont {D.~R.}\ \bibnamefont
			{Hofstadter}},\ }\bibfield  {title} {\bibinfo {title} {Energy levels and wave
			functions of bloch electrons in rational and irrational magnetic fields},\
	}\href {https://doi.org/10.1103/PhysRevB.14.2239} {\bibfield  {journal}
		{\bibinfo  {journal} {Phys. Rev. B}\ }\textbf {\bibinfo {volume} {14}},\
		\bibinfo {pages} {2239} (\bibinfo {year} {1976})}\BibitemShut {NoStop}%
	\bibitem [{\citenamefont {Goldman}\ \emph {et~al.}(2014)\citenamefont
		{Goldman}, \citenamefont {Juzeliūnas}, \citenamefont {Öhberg},\ and\
		\citenamefont {Spielman}}]{Goldman_2014}%
	\BibitemOpen
	\bibfield  {author} {\bibinfo {author} {\bibfnamefont {N.}~\bibnamefont
			{Goldman}}, \bibinfo {author} {\bibfnamefont {G.}~\bibnamefont
			{Juzeliūnas}}, \bibinfo {author} {\bibfnamefont {P.}~\bibnamefont
			{Öhberg}},\ and\ \bibinfo {author} {\bibfnamefont {I.~B.}\ \bibnamefont
			{Spielman}},\ }\bibfield  {title} {\bibinfo {title} {Light-induced gauge
			fields for ultracold atoms},\ }\href
	{https://doi.org/10.1088/0034-4885/77/12/126401} {\bibfield  {journal}
		{\bibinfo  {journal} {Reports on Progress in Physics}\ }\textbf {\bibinfo
			{volume} {77}},\ \bibinfo {pages} {126401} (\bibinfo {year}
		{2014})}\BibitemShut {NoStop}%
	\bibitem [{\citenamefont {Lindner}\ \emph {et~al.}(2011)\citenamefont
		{Lindner}, \citenamefont {Refael},\ and\ \citenamefont {Galitski}}]{RN6}%
	\BibitemOpen
	\bibfield  {author} {\bibinfo {author} {\bibfnamefont {N.~H.}\ \bibnamefont
			{Lindner}}, \bibinfo {author} {\bibfnamefont {G.}~\bibnamefont {Refael}},\
		and\ \bibinfo {author} {\bibfnamefont {V.}~\bibnamefont {Galitski}},\
	}\bibfield  {title} {\bibinfo {title} {Floquet topological insulator in
			semiconductor quantum wells},\ }\href {https://doi.org/10.1038/nphys1926}
	{\bibfield  {journal} {\bibinfo  {journal} {Nature Physics}\ }\textbf
		{\bibinfo {volume} {7}},\ \bibinfo {pages} {490–495} (\bibinfo {year}
		{2011})}\BibitemShut {NoStop}%
	\bibitem [{\citenamefont {Jotzu}\ \emph {et~al.}(2014)\citenamefont {Jotzu},
		\citenamefont {Messer}, \citenamefont {Desbuquois}, \citenamefont {Lebrat},
		\citenamefont {Uehlinger}, \citenamefont {Greif},\ and\ \citenamefont
		{Esslinger}}]{RN8}%
	\BibitemOpen
	\bibfield  {author} {\bibinfo {author} {\bibfnamefont {G.}~\bibnamefont
			{Jotzu}}, \bibinfo {author} {\bibfnamefont {M.}~\bibnamefont {Messer}},
		\bibinfo {author} {\bibfnamefont {R.}~\bibnamefont {Desbuquois}}, \bibinfo
		{author} {\bibfnamefont {M.}~\bibnamefont {Lebrat}}, \bibinfo {author}
		{\bibfnamefont {T.}~\bibnamefont {Uehlinger}}, \bibinfo {author}
		{\bibfnamefont {D.}~\bibnamefont {Greif}},\ and\ \bibinfo {author}
		{\bibfnamefont {T.}~\bibnamefont {Esslinger}},\ }\bibfield  {title} {\bibinfo
		{title} {Experimental realization of the topological haldane model with
			ultracold fermions},\ }\href {https://doi.org/10.1038/nature13915} {\bibfield
		{journal} {\bibinfo  {journal} {Nature}\ }\textbf {\bibinfo {volume}
			{515}},\ \bibinfo {pages} {237–240} (\bibinfo {year} {2014})}\BibitemShut
	{NoStop}%
	\bibitem [{\citenamefont {Ashida}\ \emph {et~al.}(2020)\citenamefont {Ashida},
		\citenamefont {Gong},\ and\ \citenamefont {Ueda}}]{Ashida02072020}%
	\BibitemOpen
	\bibfield  {author} {\bibinfo {author} {\bibfnamefont {Y.}~\bibnamefont
			{Ashida}}, \bibinfo {author} {\bibfnamefont {Z.}~\bibnamefont {Gong}},\ and\
		\bibinfo {author} {\bibfnamefont {M.}~\bibnamefont {Ueda}},\ }\bibfield
	{title} {\bibinfo {title} {Non-hermitian physics},\ }\href
	{https://doi.org/10.1080/00018732.2021.1876991} {\bibfield  {journal}
		{\bibinfo  {journal} {Advances in Physics}\ }\textbf {\bibinfo {volume}
			{69}},\ \bibinfo {pages} {249} (\bibinfo {year} {2020})}\BibitemShut
	{NoStop}%
	\bibitem [{\citenamefont {Bergholtz}\ \emph {et~al.}(2021)\citenamefont
		{Bergholtz}, \citenamefont {Budich},\ and\ \citenamefont
		{Kunst}}]{RevModPhys.93.015005}%
	\BibitemOpen
	\bibfield  {author} {\bibinfo {author} {\bibfnamefont {E.~J.}\ \bibnamefont
			{Bergholtz}}, \bibinfo {author} {\bibfnamefont {J.~C.}\ \bibnamefont
			{Budich}},\ and\ \bibinfo {author} {\bibfnamefont {F.~K.}\ \bibnamefont
			{Kunst}},\ }\bibfield  {title} {\bibinfo {title} {Exceptional topology of
			non-hermitian systems},\ }\href
	{https://doi.org/10.1103/RevModPhys.93.015005} {\bibfield  {journal}
		{\bibinfo  {journal} {Rev. Mod. Phys.}\ }\textbf {\bibinfo {volume} {93}},\
		\bibinfo {pages} {015005} (\bibinfo {year} {2021})}\BibitemShut {NoStop}%
	\bibitem [{\citenamefont {Okuma}\ and\ \citenamefont
		{Sato}(2023)}]{annurevconmatphys}%
	\BibitemOpen
	\bibfield  {author} {\bibinfo {author} {\bibfnamefont {N.}~\bibnamefont
			{Okuma}}\ and\ \bibinfo {author} {\bibfnamefont {M.}~\bibnamefont {Sato}},\
	}\bibfield  {title} {\bibinfo {title} {Non-hermitian topological phenomena: A
			review},\ }\href
	{https://doi.org/https://doi.org/10.1146/annurev-conmatphys-040521-033133}
	{\bibfield  {journal} {\bibinfo  {journal} {Annual Review of Condensed Matter
				Physics}\ }\textbf {\bibinfo {volume} {14}},\ \bibinfo {pages} {83} (\bibinfo
		{year} {2023})}\BibitemShut {NoStop}%
	\bibitem [{\citenamefont {Zhang}\ \emph {et~al.}(2022)\citenamefont {Zhang},
		\citenamefont {Zhang}, \citenamefont {Lu},\ and\ \citenamefont
		{Chen}}]{zhang2022review}%
	\BibitemOpen
	\bibfield  {author} {\bibinfo {author} {\bibfnamefont {X.}~\bibnamefont
			{Zhang}}, \bibinfo {author} {\bibfnamefont {T.}~\bibnamefont {Zhang}},
		\bibinfo {author} {\bibfnamefont {M.-H.}\ \bibnamefont {Lu}},\ and\ \bibinfo
		{author} {\bibfnamefont {Y.-F.}\ \bibnamefont {Chen}},\ }\bibfield  {title}
	{\bibinfo {title} {A review on non-hermitian skin effect},\ }\href@noop {}
	{\bibfield  {journal} {\bibinfo  {journal} {Advances in Physics: X}\ }\textbf
		{\bibinfo {volume} {7}},\ \bibinfo {pages} {2109431} (\bibinfo {year}
		{2022})}\BibitemShut {NoStop}%
	\bibitem [{\citenamefont {Lin}\ \emph {et~al.}(2023)\citenamefont {Lin},
		\citenamefont {Tai}, \citenamefont {Li},\ and\ \citenamefont {Lee}}]{RN10}%
	\BibitemOpen
	\bibfield  {author} {\bibinfo {author} {\bibfnamefont {R.}~\bibnamefont
			{Lin}}, \bibinfo {author} {\bibfnamefont {T.}~\bibnamefont {Tai}}, \bibinfo
		{author} {\bibfnamefont {L.}~\bibnamefont {Li}},\ and\ \bibinfo {author}
		{\bibfnamefont {C.~H.}\ \bibnamefont {Lee}},\ }\bibfield  {title} {\bibinfo
		{title} {Topological non-hermitian skin effect},\ }\href
	{https://doi.org/10.1007/s11467-023-1309-z} {\bibfield  {journal} {\bibinfo
			{journal} {Frontiers of Physics}\ }\textbf {\bibinfo {volume} {18}},\
		\bibinfo {pages} {53605} (\bibinfo {year} {2023})}\BibitemShut {NoStop}%
	\bibitem [{\citenamefont {{\"O}zdemir}\ \emph {et~al.}(2019)\citenamefont
		{{\"O}zdemir}, \citenamefont {Rotter}, \citenamefont {Nori},\ and\
		\citenamefont {Yang}}]{RN11}%
	\BibitemOpen
	\bibfield  {author} {\bibinfo {author} {\bibfnamefont {{\c{S}}.~K.}\
			\bibnamefont {{\"O}zdemir}}, \bibinfo {author} {\bibfnamefont
			{S.}~\bibnamefont {Rotter}}, \bibinfo {author} {\bibfnamefont
			{F.}~\bibnamefont {Nori}},\ and\ \bibinfo {author} {\bibfnamefont
			{L.}~\bibnamefont {Yang}},\ }\bibfield  {title} {\bibinfo {title}
		{Parity–time symmetry and exceptional points in photonics},\ }\href
	{https://doi.org/10.1038/s41563-019-0304-9} {\bibfield  {journal} {\bibinfo
			{journal} {Nature Materials}\ }\textbf {\bibinfo {volume} {18}},\ \bibinfo
		{pages} {783–798} (\bibinfo {year} {2019})}\BibitemShut {NoStop}%
	\bibitem [{\citenamefont {Gong}\ \emph {et~al.}(2018)\citenamefont {Gong},
		\citenamefont {Ashida}, \citenamefont {Kawabata}, \citenamefont {Takasan},
		\citenamefont {Higashikawa},\ and\ \citenamefont {Ueda}}]{PhysRevX.8.031079}%
	\BibitemOpen
	\bibfield  {author} {\bibinfo {author} {\bibfnamefont {Z.}~\bibnamefont
			{Gong}}, \bibinfo {author} {\bibfnamefont {Y.}~\bibnamefont {Ashida}},
		\bibinfo {author} {\bibfnamefont {K.}~\bibnamefont {Kawabata}}, \bibinfo
		{author} {\bibfnamefont {K.}~\bibnamefont {Takasan}}, \bibinfo {author}
		{\bibfnamefont {S.}~\bibnamefont {Higashikawa}},\ and\ \bibinfo {author}
		{\bibfnamefont {M.}~\bibnamefont {Ueda}},\ }\bibfield  {title} {\bibinfo
		{title} {Topological phases of non-hermitian systems},\ }\href
	{https://doi.org/10.1103/PhysRevX.8.031079} {\bibfield  {journal} {\bibinfo
			{journal} {Phys. Rev. X}\ }\textbf {\bibinfo {volume} {8}},\ \bibinfo {pages}
		{031079} (\bibinfo {year} {2018})}\BibitemShut {NoStop}%
	\bibitem [{\citenamefont {Shen}\ \emph {et~al.}(2018)\citenamefont {Shen},
		\citenamefont {Zhen},\ and\ \citenamefont {Fu}}]{PhysRevLett.120.146402}%
	\BibitemOpen
	\bibfield  {author} {\bibinfo {author} {\bibfnamefont {H.}~\bibnamefont
			{Shen}}, \bibinfo {author} {\bibfnamefont {B.}~\bibnamefont {Zhen}},\ and\
		\bibinfo {author} {\bibfnamefont {L.}~\bibnamefont {Fu}},\ }\bibfield
	{title} {\bibinfo {title} {Topological band theory for non-hermitian
			hamiltonians},\ }\href {https://doi.org/10.1103/PhysRevLett.120.146402}
	{\bibfield  {journal} {\bibinfo  {journal} {Phys. Rev. Lett.}\ }\textbf
		{\bibinfo {volume} {120}},\ \bibinfo {pages} {146402} (\bibinfo {year}
		{2018})}\BibitemShut {NoStop}%
	\bibitem [{\citenamefont {Kawabata}\ \emph {et~al.}(2019)\citenamefont
		{Kawabata}, \citenamefont {Shiozaki}, \citenamefont {Ueda},\ and\
		\citenamefont {Sato}}]{PhysRevX.9.041015}%
	\BibitemOpen
	\bibfield  {author} {\bibinfo {author} {\bibfnamefont {K.}~\bibnamefont
			{Kawabata}}, \bibinfo {author} {\bibfnamefont {K.}~\bibnamefont {Shiozaki}},
		\bibinfo {author} {\bibfnamefont {M.}~\bibnamefont {Ueda}},\ and\ \bibinfo
		{author} {\bibfnamefont {M.}~\bibnamefont {Sato}},\ }\bibfield  {title}
	{\bibinfo {title} {Symmetry and topology in non-hermitian physics},\ }\href
	{https://doi.org/10.1103/PhysRevX.9.041015} {\bibfield  {journal} {\bibinfo
			{journal} {Phys. Rev. X}\ }\textbf {\bibinfo {volume} {9}},\ \bibinfo {pages}
		{041015} (\bibinfo {year} {2019})}\BibitemShut {NoStop}%
	\bibitem [{\citenamefont {Yao}\ and\ \citenamefont
		{Wang}(2018)}]{PhysRevLett.121.086803}%
	\BibitemOpen
	\bibfield  {author} {\bibinfo {author} {\bibfnamefont {S.}~\bibnamefont
			{Yao}}\ and\ \bibinfo {author} {\bibfnamefont {Z.}~\bibnamefont {Wang}},\
	}\bibfield  {title} {\bibinfo {title} {Edge states and topological invariants
			of non-hermitian systems},\ }\href
	{https://doi.org/10.1103/PhysRevLett.121.086803} {\bibfield  {journal}
		{\bibinfo  {journal} {Phys. Rev. Lett.}\ }\textbf {\bibinfo {volume} {121}},\
		\bibinfo {pages} {086803} (\bibinfo {year} {2018})}\BibitemShut {NoStop}%
	\bibitem [{\citenamefont {Kunst}\ \emph {et~al.}(2018)\citenamefont {Kunst},
		\citenamefont {Edvardsson}, \citenamefont {Budich},\ and\ \citenamefont
		{Bergholtz}}]{PhysRevLett.121.026808}%
	\BibitemOpen
	\bibfield  {author} {\bibinfo {author} {\bibfnamefont {F.~K.}\ \bibnamefont
			{Kunst}}, \bibinfo {author} {\bibfnamefont {E.}~\bibnamefont {Edvardsson}},
		\bibinfo {author} {\bibfnamefont {J.~C.}\ \bibnamefont {Budich}},\ and\
		\bibinfo {author} {\bibfnamefont {E.~J.}\ \bibnamefont {Bergholtz}},\
	}\bibfield  {title} {\bibinfo {title} {Biorthogonal bulk-boundary
			correspondence in non-hermitian systems},\ }\href
	{https://doi.org/10.1103/PhysRevLett.121.026808} {\bibfield  {journal}
		{\bibinfo  {journal} {Phys. Rev. Lett.}\ }\textbf {\bibinfo {volume} {121}},\
		\bibinfo {pages} {026808} (\bibinfo {year} {2018})}\BibitemShut {NoStop}%
	\bibitem [{\citenamefont {Okuma}\ \emph {et~al.}(2020)\citenamefont {Okuma},
		\citenamefont {Kawabata}, \citenamefont {Shiozaki},\ and\ \citenamefont
		{Sato}}]{PhysRevLett.124.086801}%
	\BibitemOpen
	\bibfield  {author} {\bibinfo {author} {\bibfnamefont {N.}~\bibnamefont
			{Okuma}}, \bibinfo {author} {\bibfnamefont {K.}~\bibnamefont {Kawabata}},
		\bibinfo {author} {\bibfnamefont {K.}~\bibnamefont {Shiozaki}},\ and\
		\bibinfo {author} {\bibfnamefont {M.}~\bibnamefont {Sato}},\ }\bibfield
	{title} {\bibinfo {title} {Topological origin of non-hermitian skin
			effects},\ }\href {https://doi.org/10.1103/PhysRevLett.124.086801} {\bibfield
		{journal} {\bibinfo  {journal} {Phys. Rev. Lett.}\ }\textbf {\bibinfo
			{volume} {124}},\ \bibinfo {pages} {086801} (\bibinfo {year}
		{2020})}\BibitemShut {NoStop}%
	\bibitem [{\citenamefont {Lee}\ and\ \citenamefont
		{Thomale}(2019)}]{PhysRevB.99.201103}%
	\BibitemOpen
	\bibfield  {author} {\bibinfo {author} {\bibfnamefont {C.~H.}\ \bibnamefont
			{Lee}}\ and\ \bibinfo {author} {\bibfnamefont {R.}~\bibnamefont {Thomale}},\
	}\bibfield  {title} {\bibinfo {title} {Anatomy of skin modes and topology in
			non-hermitian systems},\ }\href {https://doi.org/10.1103/PhysRevB.99.201103}
	{\bibfield  {journal} {\bibinfo  {journal} {Phys. Rev. B}\ }\textbf {\bibinfo
			{volume} {99}},\ \bibinfo {pages} {201103(R)} (\bibinfo {year}
		{2019})}\BibitemShut {NoStop}%
	\bibitem [{\citenamefont {Hu}\ and\ \citenamefont
		{Zhao}(2021)}]{PhysRevLett.126.010401}%
	\BibitemOpen
	\bibfield  {author} {\bibinfo {author} {\bibfnamefont {H.}~\bibnamefont
			{Hu}}\ and\ \bibinfo {author} {\bibfnamefont {E.}~\bibnamefont {Zhao}},\
	}\bibfield  {title} {\bibinfo {title} {Knots and non-hermitian bloch bands},\
	}\href {https://doi.org/10.1103/PhysRevLett.126.010401} {\bibfield  {journal}
		{\bibinfo  {journal} {Phys. Rev. Lett.}\ }\textbf {\bibinfo {volume} {126}},\
		\bibinfo {pages} {010401} (\bibinfo {year} {2021})}\BibitemShut {NoStop}%
	\bibitem [{\citenamefont {Yao}\ \emph {et~al.}(2018)\citenamefont {Yao},
		\citenamefont {Song},\ and\ \citenamefont {Wang}}]{PhysRevLett.121.136802}%
	\BibitemOpen
	\bibfield  {author} {\bibinfo {author} {\bibfnamefont {S.}~\bibnamefont
			{Yao}}, \bibinfo {author} {\bibfnamefont {F.}~\bibnamefont {Song}},\ and\
		\bibinfo {author} {\bibfnamefont {Z.}~\bibnamefont {Wang}},\ }\bibfield
	{title} {\bibinfo {title} {Non-hermitian chern bands},\ }\href
	{https://doi.org/10.1103/PhysRevLett.121.136802} {\bibfield  {journal}
		{\bibinfo  {journal} {Phys. Rev. Lett.}\ }\textbf {\bibinfo {volume} {121}},\
		\bibinfo {pages} {136802} (\bibinfo {year} {2018})}\BibitemShut {NoStop}%
	\bibitem [{\citenamefont {Kawabata}\ \emph {et~al.}(2018)\citenamefont
		{Kawabata}, \citenamefont {Shiozaki},\ and\ \citenamefont
		{Ueda}}]{PhysRevB.98.165148}%
	\BibitemOpen
	\bibfield  {author} {\bibinfo {author} {\bibfnamefont {K.}~\bibnamefont
			{Kawabata}}, \bibinfo {author} {\bibfnamefont {K.}~\bibnamefont {Shiozaki}},\
		and\ \bibinfo {author} {\bibfnamefont {M.}~\bibnamefont {Ueda}},\ }\bibfield
	{title} {\bibinfo {title} {Anomalous helical edge states in a non-hermitian
			chern insulator},\ }\href {https://doi.org/10.1103/PhysRevB.98.165148}
	{\bibfield  {journal} {\bibinfo  {journal} {Phys. Rev. B}\ }\textbf {\bibinfo
			{volume} {98}},\ \bibinfo {pages} {165148} (\bibinfo {year}
		{2018})}\BibitemShut {NoStop}%
	\bibitem [{\citenamefont {Wu}\ \emph {et~al.}(2019)\citenamefont {Wu},
		\citenamefont {Jin},\ and\ \citenamefont {Song}}]{PhysRevB.100.155117}%
	\BibitemOpen
	\bibfield  {author} {\bibinfo {author} {\bibfnamefont {H.~C.}\ \bibnamefont
			{Wu}}, \bibinfo {author} {\bibfnamefont {L.}~\bibnamefont {Jin}},\ and\
		\bibinfo {author} {\bibfnamefont {Z.}~\bibnamefont {Song}},\ }\bibfield
	{title} {\bibinfo {title} {Inversion symmetric non-hermitian chern
			insulator},\ }\href {https://doi.org/10.1103/PhysRevB.100.155117} {\bibfield
		{journal} {\bibinfo  {journal} {Phys. Rev. B}\ }\textbf {\bibinfo {volume}
			{100}},\ \bibinfo {pages} {155117} (\bibinfo {year} {2019})}\BibitemShut
	{NoStop}%
	\bibitem [{\citenamefont {Hirsbrunner}\ \emph {et~al.}(2019)\citenamefont
		{Hirsbrunner}, \citenamefont {Philip},\ and\ \citenamefont
		{Gilbert}}]{PhysRevB.100.081104}%
	\BibitemOpen
	\bibfield  {author} {\bibinfo {author} {\bibfnamefont {M.~R.}\ \bibnamefont
			{Hirsbrunner}}, \bibinfo {author} {\bibfnamefont {T.~M.}\ \bibnamefont
			{Philip}},\ and\ \bibinfo {author} {\bibfnamefont {M.~J.}\ \bibnamefont
			{Gilbert}},\ }\bibfield  {title} {\bibinfo {title} {Topology and observables
			of the non-hermitian chern insulator},\ }\href
	{https://doi.org/10.1103/PhysRevB.100.081104} {\bibfield  {journal} {\bibinfo
			{journal} {Phys. Rev. B}\ }\textbf {\bibinfo {volume} {100}},\ \bibinfo
		{pages} {081104(R)} (\bibinfo {year} {2019})}\BibitemShut {NoStop}%
	\bibitem [{\citenamefont {Lee}\ \emph {et~al.}(2019)\citenamefont {Lee},
		\citenamefont {Li},\ and\ \citenamefont {Gong}}]{PhysRevLett.123.016805}%
	\BibitemOpen
	\bibfield  {author} {\bibinfo {author} {\bibfnamefont {C.~H.}\ \bibnamefont
			{Lee}}, \bibinfo {author} {\bibfnamefont {L.}~\bibnamefont {Li}},\ and\
		\bibinfo {author} {\bibfnamefont {J.}~\bibnamefont {Gong}},\ }\bibfield
	{title} {\bibinfo {title} {Hybrid higher-order skin-topological modes in
			nonreciprocal systems},\ }\href
	{https://doi.org/10.1103/PhysRevLett.123.016805} {\bibfield  {journal}
		{\bibinfo  {journal} {Phys. Rev. Lett.}\ }\textbf {\bibinfo {volume} {123}},\
		\bibinfo {pages} {016805} (\bibinfo {year} {2019})}\BibitemShut {NoStop}%
	\bibitem [{\citenamefont {Zhu}\ and\ \citenamefont
		{Gong}(2022)}]{PhysRevB.106.035425}%
	\BibitemOpen
	\bibfield  {author} {\bibinfo {author} {\bibfnamefont {W.}~\bibnamefont
			{Zhu}}\ and\ \bibinfo {author} {\bibfnamefont {J.}~\bibnamefont {Gong}},\
	}\bibfield  {title} {\bibinfo {title} {Hybrid skin-topological modes without
			asymmetric couplings},\ }\href {https://doi.org/10.1103/PhysRevB.106.035425}
	{\bibfield  {journal} {\bibinfo  {journal} {Phys. Rev. B}\ }\textbf {\bibinfo
			{volume} {106}},\ \bibinfo {pages} {035425} (\bibinfo {year}
		{2022})}\BibitemShut {NoStop}%
	\bibitem [{\citenamefont {Panigrahi}\ \emph {et~al.}(2022)\citenamefont
		{Panigrahi}, \citenamefont {Moessner},\ and\ \citenamefont
		{Roy}}]{PhysRevB.106.L041302}%
	\BibitemOpen
	\bibfield  {author} {\bibinfo {author} {\bibfnamefont {A.}~\bibnamefont
			{Panigrahi}}, \bibinfo {author} {\bibfnamefont {R.}~\bibnamefont
			{Moessner}},\ and\ \bibinfo {author} {\bibfnamefont {B.}~\bibnamefont
			{Roy}},\ }\bibfield  {title} {\bibinfo {title} {Non-hermitian dislocation
			modes: Stability and melting across exceptional points},\ }\href
	{https://doi.org/10.1103/PhysRevB.106.L041302} {\bibfield  {journal}
		{\bibinfo  {journal} {Phys. Rev. B}\ }\textbf {\bibinfo {volume} {106}},\
		\bibinfo {pages} {L041302} (\bibinfo {year} {2022})}\BibitemShut {NoStop}%
	\bibitem [{\citenamefont {Liu}\ \emph {et~al.}(2024)\citenamefont {Liu},
		\citenamefont {Mandal}, \citenamefont {Zhou}, \citenamefont {Xi},
		\citenamefont {Banerjee}, \citenamefont {Hu}, \citenamefont {Wei},
		\citenamefont {Wang}, \citenamefont {Wang}, \citenamefont {Gao},
		\citenamefont {Chen}, \citenamefont {Yang}, \citenamefont {Chong},\ and\
		\citenamefont {Zhang}}]{PhysRevLett.132.113802}%
	\BibitemOpen
	\bibfield  {author} {\bibinfo {author} {\bibfnamefont {G.-G.}\ \bibnamefont
			{Liu}}, \bibinfo {author} {\bibfnamefont {S.}~\bibnamefont {Mandal}},
		\bibinfo {author} {\bibfnamefont {P.}~\bibnamefont {Zhou}}, \bibinfo {author}
		{\bibfnamefont {X.}~\bibnamefont {Xi}}, \bibinfo {author} {\bibfnamefont
			{R.}~\bibnamefont {Banerjee}}, \bibinfo {author} {\bibfnamefont {Y.-H.}\
			\bibnamefont {Hu}}, \bibinfo {author} {\bibfnamefont {M.}~\bibnamefont
			{Wei}}, \bibinfo {author} {\bibfnamefont {M.}~\bibnamefont {Wang}}, \bibinfo
		{author} {\bibfnamefont {Q.}~\bibnamefont {Wang}}, \bibinfo {author}
		{\bibfnamefont {Z.}~\bibnamefont {Gao}}, \bibinfo {author} {\bibfnamefont
			{H.}~\bibnamefont {Chen}}, \bibinfo {author} {\bibfnamefont {Y.}~\bibnamefont
			{Yang}}, \bibinfo {author} {\bibfnamefont {Y.}~\bibnamefont {Chong}},\ and\
		\bibinfo {author} {\bibfnamefont {B.}~\bibnamefont {Zhang}},\ }\bibfield
	{title} {\bibinfo {title} {Localization of chiral edge states by the
			non-hermitian skin effect},\ }\href
	{https://doi.org/10.1103/PhysRevLett.132.113802} {\bibfield  {journal}
		{\bibinfo  {journal} {Phys. Rev. Lett.}\ }\textbf {\bibinfo {volume} {132}},\
		\bibinfo {pages} {113802} (\bibinfo {year} {2024})}\BibitemShut {NoStop}%
	\bibitem [{\citenamefont {Sun}\ \emph {et~al.}(2024)\citenamefont {Sun},
		\citenamefont {Hou}, \citenamefont {Wan}, \citenamefont {Wang}, \citenamefont
		{Zhu}, \citenamefont {Ruan},\ and\ \citenamefont
		{Yang}}]{PhysRevLett.132.063804}%
	\BibitemOpen
	\bibfield  {author} {\bibinfo {author} {\bibfnamefont {Y.}~\bibnamefont
			{Sun}}, \bibinfo {author} {\bibfnamefont {X.}~\bibnamefont {Hou}}, \bibinfo
		{author} {\bibfnamefont {T.}~\bibnamefont {Wan}}, \bibinfo {author}
		{\bibfnamefont {F.}~\bibnamefont {Wang}}, \bibinfo {author} {\bibfnamefont
			{S.}~\bibnamefont {Zhu}}, \bibinfo {author} {\bibfnamefont {Z.}~\bibnamefont
			{Ruan}},\ and\ \bibinfo {author} {\bibfnamefont {Z.}~\bibnamefont {Yang}},\
	}\bibfield  {title} {\bibinfo {title} {Photonic floquet skin-topological
			effect},\ }\href {https://doi.org/10.1103/PhysRevLett.132.063804} {\bibfield
		{journal} {\bibinfo  {journal} {Phys. Rev. Lett.}\ }\textbf {\bibinfo
			{volume} {132}},\ \bibinfo {pages} {063804} (\bibinfo {year}
		{2024})}\BibitemShut {NoStop}%
	\bibitem [{\citenamefont {Wu}\ \emph {et~al.}(2024)\citenamefont {Wu},
		\citenamefont {Zheng}, \citenamefont {Liang}, \citenamefont {Ke},
		\citenamefont {Lu}, \citenamefont {Deng}, \citenamefont {Huang},\ and\
		\citenamefont {Liu}}]{PhysRevLett.133.126601}%
	\BibitemOpen
	\bibfield  {author} {\bibinfo {author} {\bibfnamefont {J.}~\bibnamefont
			{Wu}}, \bibinfo {author} {\bibfnamefont {R.}~\bibnamefont {Zheng}}, \bibinfo
		{author} {\bibfnamefont {J.}~\bibnamefont {Liang}}, \bibinfo {author}
		{\bibfnamefont {M.}~\bibnamefont {Ke}}, \bibinfo {author} {\bibfnamefont
			{J.}~\bibnamefont {Lu}}, \bibinfo {author} {\bibfnamefont {W.}~\bibnamefont
			{Deng}}, \bibinfo {author} {\bibfnamefont {X.}~\bibnamefont {Huang}},\ and\
		\bibinfo {author} {\bibfnamefont {Z.}~\bibnamefont {Liu}},\ }\bibfield
	{title} {\bibinfo {title} {Spin-dependent localization of helical edge states
			in a non-hermitian phononic crystal},\ }\href
	{https://doi.org/10.1103/PhysRevLett.133.126601} {\bibfield  {journal}
		{\bibinfo  {journal} {Phys. Rev. Lett.}\ }\textbf {\bibinfo {volume} {133}},\
		\bibinfo {pages} {126601} (\bibinfo {year} {2024})}\BibitemShut {NoStop}%
	\bibitem [{\citenamefont {Metelmann}\ and\ \citenamefont
		{Clerk}(2015)}]{PhysRevX.5.021025}%
	\BibitemOpen
	\bibfield  {author} {\bibinfo {author} {\bibfnamefont {A.}~\bibnamefont
			{Metelmann}}\ and\ \bibinfo {author} {\bibfnamefont {A.~A.}\ \bibnamefont
			{Clerk}},\ }\bibfield  {title} {\bibinfo {title} {Nonreciprocal photon
			transmission and amplification via reservoir engineering},\ }\href
	{https://doi.org/10.1103/PhysRevX.5.021025} {\bibfield  {journal} {\bibinfo
			{journal} {Phys. Rev. X}\ }\textbf {\bibinfo {volume} {5}},\ \bibinfo {pages}
		{021025} (\bibinfo {year} {2015})}\BibitemShut {NoStop}%
	\bibitem [{\citenamefont {Gao}\ \emph {et~al.}(2021)\citenamefont {Gao},
		\citenamefont {Xue}, \citenamefont {Gu}, \citenamefont {Liu}, \citenamefont
		{Zhu},\ and\ \citenamefont {Zhang}}]{RN12}%
	\BibitemOpen
	\bibfield  {author} {\bibinfo {author} {\bibfnamefont {H.}~\bibnamefont
			{Gao}}, \bibinfo {author} {\bibfnamefont {H.}~\bibnamefont {Xue}}, \bibinfo
		{author} {\bibfnamefont {Z.}~\bibnamefont {Gu}}, \bibinfo {author}
		{\bibfnamefont {T.}~\bibnamefont {Liu}}, \bibinfo {author} {\bibfnamefont
			{J.}~\bibnamefont {Zhu}},\ and\ \bibinfo {author} {\bibfnamefont
			{B.}~\bibnamefont {Zhang}},\ }\bibfield  {title} {\bibinfo {title}
		{Non-hermitian route to higher-order topology in an acoustic crystal},\
	}\href {https://doi.org/10.1038/s41467-021-22223-y} {\bibfield  {journal}
		{\bibinfo  {journal} {Nature Communications}\ }\textbf {\bibinfo {volume}
			{12}},\ \bibinfo {pages} {1888} (\bibinfo {year} {2021})}\BibitemShut
	{NoStop}%
	\bibitem [{\citenamefont {Zhang}\ \emph {et~al.}(2021)\citenamefont {Zhang},
		\citenamefont {Tian}, \citenamefont {Jiang}, \citenamefont {Lu},\ and\
		\citenamefont {Chen}}]{RN13}%
	\BibitemOpen
	\bibfield  {author} {\bibinfo {author} {\bibfnamefont {X.}~\bibnamefont
			{Zhang}}, \bibinfo {author} {\bibfnamefont {Y.}~\bibnamefont {Tian}},
		\bibinfo {author} {\bibfnamefont {J.-H.}\ \bibnamefont {Jiang}}, \bibinfo
		{author} {\bibfnamefont {M.-H.}\ \bibnamefont {Lu}},\ and\ \bibinfo {author}
		{\bibfnamefont {Y.-F.}\ \bibnamefont {Chen}},\ }\bibfield  {title} {\bibinfo
		{title} {Observation of higher-order non-hermitian skin effect},\ }\href
	{https://doi.org/10.1038/s41467-021-25716-y} {\bibfield  {journal} {\bibinfo
			{journal} {Nature Communications}\ }\textbf {\bibinfo {volume} {12}},\
		\bibinfo {pages} {5377} (\bibinfo {year} {2021})}\BibitemShut {NoStop}%
	\bibitem [{\citenamefont {Wang}\ \emph {et~al.}(2022)\citenamefont {Wang},
		\citenamefont {Wang},\ and\ \citenamefont {Ma}}]{RN14}%
	\BibitemOpen
	\bibfield  {author} {\bibinfo {author} {\bibfnamefont {W.}~\bibnamefont
			{Wang}}, \bibinfo {author} {\bibfnamefont {X.}~\bibnamefont {Wang}},\ and\
		\bibinfo {author} {\bibfnamefont {G.}~\bibnamefont {Ma}},\ }\bibfield
	{title} {\bibinfo {title} {Non-hermitian morphing of topological modes},\
	}\href {https://doi.org/10.1038/s41586-022-04929-1} {\bibfield  {journal}
		{\bibinfo  {journal} {Nature}\ }\textbf {\bibinfo {volume} {608}},\ \bibinfo
		{pages} {50–55} (\bibinfo {year} {2022})}\BibitemShut {NoStop}%
	\bibitem [{\citenamefont {Huang}\ \emph {et~al.}(2024)\citenamefont {Huang},
		\citenamefont {Huang}, \citenamefont {Shen}, \citenamefont {Yves},
		\citenamefont {Pilipchuk}, \citenamefont {Ni}, \citenamefont {Kim},
		\citenamefont {Chiang}, \citenamefont {Powell}, \citenamefont {Zhu},
		\citenamefont {Cheng}, \citenamefont {Li}, \citenamefont {Sadreev},
		\citenamefont {Alù},\ and\ \citenamefont {Miroshnichenko}}]{RN15}%
	\BibitemOpen
	\bibfield  {author} {\bibinfo {author} {\bibfnamefont {L.}~\bibnamefont
			{Huang}}, \bibinfo {author} {\bibfnamefont {S.}~\bibnamefont {Huang}},
		\bibinfo {author} {\bibfnamefont {C.}~\bibnamefont {Shen}}, \bibinfo {author}
		{\bibfnamefont {S.}~\bibnamefont {Yves}}, \bibinfo {author} {\bibfnamefont
			{A.~S.}\ \bibnamefont {Pilipchuk}}, \bibinfo {author} {\bibfnamefont
			{X.}~\bibnamefont {Ni}}, \bibinfo {author} {\bibfnamefont {S.}~\bibnamefont
			{Kim}}, \bibinfo {author} {\bibfnamefont {Y.~K.}\ \bibnamefont {Chiang}},
		\bibinfo {author} {\bibfnamefont {D.~A.}\ \bibnamefont {Powell}}, \bibinfo
		{author} {\bibfnamefont {J.}~\bibnamefont {Zhu}}, \bibinfo {author}
		{\bibfnamefont {Y.}~\bibnamefont {Cheng}}, \bibinfo {author} {\bibfnamefont
			{Y.}~\bibnamefont {Li}}, \bibinfo {author} {\bibfnamefont {A.~F.}\
			\bibnamefont {Sadreev}}, \bibinfo {author} {\bibfnamefont {A.}~\bibnamefont
			{Alù}},\ and\ \bibinfo {author} {\bibfnamefont {A.~E.}\ \bibnamefont
			{Miroshnichenko}},\ }\bibfield  {title} {\bibinfo {title} {Acoustic
			resonances in non-hermitian open systems},\ }\href
	{https://doi.org/10.1038/s42254-023-00659-z} {\bibfield  {journal} {\bibinfo
			{journal} {Nature Reviews Physics}\ }\textbf {\bibinfo {volume} {6}},\
		\bibinfo {pages} {11–27} (\bibinfo {year} {2024})}\BibitemShut {NoStop}%
	\bibitem [{\citenamefont {Roccati}\ \emph {et~al.}(2024)\citenamefont
		{Roccati}, \citenamefont {Bello}, \citenamefont {Gong}, \citenamefont {Ueda},
		\citenamefont {Ciccarello}, \citenamefont {Chenu},\ and\ \citenamefont
		{Carollo}}]{RN16}%
	\BibitemOpen
	\bibfield  {author} {\bibinfo {author} {\bibfnamefont {F.}~\bibnamefont
			{Roccati}}, \bibinfo {author} {\bibfnamefont {M.}~\bibnamefont {Bello}},
		\bibinfo {author} {\bibfnamefont {Z.}~\bibnamefont {Gong}}, \bibinfo {author}
		{\bibfnamefont {M.}~\bibnamefont {Ueda}}, \bibinfo {author} {\bibfnamefont
			{F.}~\bibnamefont {Ciccarello}}, \bibinfo {author} {\bibfnamefont
			{A.}~\bibnamefont {Chenu}},\ and\ \bibinfo {author} {\bibfnamefont
			{A.}~\bibnamefont {Carollo}},\ }\bibfield  {title} {\bibinfo {title}
		{Hermitian and non-hermitian topology from photon-mediated interactions},\
	}\href {https://doi.org/10.1038/s41467-024-46471-w} {\bibfield  {journal}
		{\bibinfo  {journal} {Nature Communications}\ }\textbf {\bibinfo {volume}
			{15}},\ \bibinfo {pages} {2400} (\bibinfo {year} {2024})}\BibitemShut
	{NoStop}%
	\bibitem [{\citenamefont {Ghatak}\ \emph {et~al.}(2020)\citenamefont {Ghatak},
		\citenamefont {Brandenbourger}, \citenamefont {van Wezel},\ and\
		\citenamefont {Coulais}}]{pnas2010580117}%
	\BibitemOpen
	\bibfield  {author} {\bibinfo {author} {\bibfnamefont {A.}~\bibnamefont
			{Ghatak}}, \bibinfo {author} {\bibfnamefont {M.}~\bibnamefont
			{Brandenbourger}}, \bibinfo {author} {\bibfnamefont {J.}~\bibnamefont {van
				Wezel}},\ and\ \bibinfo {author} {\bibfnamefont {C.}~\bibnamefont
			{Coulais}},\ }\bibfield  {title} {\bibinfo {title} {Observation of
			non-hermitian topology and its bulk–edge correspondence in an active
			mechanical metamaterial},\ }\href {https://doi.org/10.1073/pnas.2010580117}
	{\bibfield  {journal} {\bibinfo  {journal} {Proceedings of the National
				Academy of Sciences}\ }\textbf {\bibinfo {volume} {117}},\ \bibinfo {pages}
		{29561} (\bibinfo {year} {2020})}\BibitemShut {NoStop}%
	\bibitem [{\citenamefont {Liu}\ \emph {et~al.}(2021)\citenamefont {Liu},
		\citenamefont {Shao}, \citenamefont {Ma}, \citenamefont {Zhang},
		\citenamefont {You}, \citenamefont {Wu}, \citenamefont {Xiang}, \citenamefont
		{Cui},\ and\ \citenamefont {Zhang}}]{5608038}%
	\BibitemOpen
	\bibfield  {author} {\bibinfo {author} {\bibfnamefont {S.}~\bibnamefont
			{Liu}}, \bibinfo {author} {\bibfnamefont {R.}~\bibnamefont {Shao}}, \bibinfo
		{author} {\bibfnamefont {S.}~\bibnamefont {Ma}}, \bibinfo {author}
		{\bibfnamefont {L.}~\bibnamefont {Zhang}}, \bibinfo {author} {\bibfnamefont
			{O.}~\bibnamefont {You}}, \bibinfo {author} {\bibfnamefont {H.}~\bibnamefont
			{Wu}}, \bibinfo {author} {\bibfnamefont {Y.~J.}\ \bibnamefont {Xiang}},
		\bibinfo {author} {\bibfnamefont {T.~J.}\ \bibnamefont {Cui}},\ and\ \bibinfo
		{author} {\bibfnamefont {S.}~\bibnamefont {Zhang}},\ }\bibfield  {title}
	{\bibinfo {title} {Non-hermitian skin effect in a non-hermitian electrical
			circuit},\ }\href {https://doi.org/10.34133/2021/5608038} {\bibfield
		{journal} {\bibinfo  {journal} {Research}\ }\textbf {\bibinfo {volume}
			{2021}},\ \bibinfo {pages} {5608038} (\bibinfo {year} {2021})}\BibitemShut
	{NoStop}%
	\bibitem [{\citenamefont {Zou}\ \emph {et~al.}(2021)\citenamefont {Zou},
		\citenamefont {Chen}, \citenamefont {He}, \citenamefont {Bao}, \citenamefont
		{Lee}, \citenamefont {Sun},\ and\ \citenamefont {Zhang}}]{RN17}%
	\BibitemOpen
	\bibfield  {author} {\bibinfo {author} {\bibfnamefont {D.}~\bibnamefont
			{Zou}}, \bibinfo {author} {\bibfnamefont {T.}~\bibnamefont {Chen}}, \bibinfo
		{author} {\bibfnamefont {W.}~\bibnamefont {He}}, \bibinfo {author}
		{\bibfnamefont {J.}~\bibnamefont {Bao}}, \bibinfo {author} {\bibfnamefont
			{C.~H.}\ \bibnamefont {Lee}}, \bibinfo {author} {\bibfnamefont
			{H.}~\bibnamefont {Sun}},\ and\ \bibinfo {author} {\bibfnamefont
			{X.}~\bibnamefont {Zhang}},\ }\bibfield  {title} {\bibinfo {title}
		{Observation of hybrid higher-order skin-topological effect in non-hermitian
			topolectrical circuits},\ }\href {https://doi.org/10.1038/s41467-021-26414-5}
	{\bibfield  {journal} {\bibinfo  {journal} {Nature Communications}\ }\textbf
		{\bibinfo {volume} {12}},\ \bibinfo {pages} {7201} (\bibinfo {year}
		{2021})}\BibitemShut {NoStop}%
\end{thebibliography}
\end{document}